\documentclass[prd,twocolumn,floats,floatfix,nofootinbib,showkeys,showpacs]{revtex4-1}
\usepackage[dvips]{graphicx}
\usepackage{graphics}
\usepackage{dcolumn}
\usepackage{amssymb}
\usepackage{bm}
\usepackage{amsmath}
\usepackage{color}

\def\spose#1{\hbox to 0pt{#1\hss}}

\def\lta{\mathrel{\spose{\lower 3pt\hbox{$\mathchar"218$}}
     \raise 2.0pt\hbox{$\mathchar"13C$}}}
\def\gta{\mathrel{\spose{\lower 3pt\hbox{$\mathchar"218$}}
     \raise 2.0pt\hbox{$\mathchar"13E$}}}
\newcommand{\be}{\begin{equation}}
\newcommand{\en}{\end{equation}}
\newcommand{\bea}{\begin{eqnarray}}
\newcommand{\ena}{\end{eqnarray}}

\graphicspath{{./figures/}}

\begin{document}

\title{Bounded Scalar Perturbations in Bouncing Brane World Cosmologies}

\author{Rodrigo Maier$^{1,2}$, Francesco Pace$^{1}$ and Ivano Dami\~ao Soares$^{3}$}

\affiliation{$^{1}$Institute of Cosmology and Gravitation, University of Portsmouth,\\
Dennis Sciama Building, Portsmouth, PO1 3FX, United Kingdom} 

\affiliation{$^{2}$ICRA - Centro Brasileiro de
Pesquisas F\'{\i}sicas -- CBPF, \\ Rua Dr. Xavier Sigaud, 150, Urca,
CEP22290-180, Rio de Janeiro, Brazil} 

\affiliation{$^{3}$Centro Brasileiro de
Pesquisas F\'{\i}sicas -- CBPF, \\ Rua Dr. Xavier Sigaud, 150, Urca,
CEP22290-180, Rio de Janeiro, Brazil}

\date{\today}

\begin{abstract}
We examine the dynamics of scalar perturbations in closed Friedmann-Lema\^itre-Robertson-Walker (FLRW)
universes in the framework of Brane World theory with a timelike extra dimension. In this scenario, the unperturbed
Friedmann equations contain additional terms arising from the bulk-brane interaction
that implement non-singular bounces in the models with a cosmological constant and non-interacting perfect fluids.
The structure of the phase-space of the models allows for two basic configurations, namely, one bounce solutions
or eternal universes. Assuming that the matter content of the model is given by dust and radiation, we derive the dynamical field
equations
for scalar hydrodynamical perturbations considering either a conformally flat (de Sitter) bulk or a perturbed bulk.
The dynamical system built with these equations is extremely involved. Nevertheless, in this paper we perform a numerical
analysis which
can shed some light on the study of cosmological scalar perturbations in bouncing brane world models. From a mathematical point
of view we show that
although the bounce enhances the amplitudes of scalar perturbations for one bounce models in the case of a de Sitter bulk,
the amplitudes of the perturbations remain sufficiently small and bounded
relative to the background values up to a certain scale.
For one bounce models in the case of a perturbed bulk
the amplitudes of all perturbations (apart from the Weyl fluid energy density) remain sufficiently small and bounded
relative to the background values for
any scale of the perturbations. We also discuss and compare the stability and bounded behaviour of the perturbations in the late
accelerated phase of one bounce solutions.
For eternal universes we argue that some of these features are maintained only for early times (typically of the order of the
first bounce).
In this sense we show that eternal solutions are highly unstable configurations considering the background model of this paper.
\end{abstract}
\keywords{Cosmology; Strings and branes}
\pacs{11.25.-w; 11.25.Db; 98.80.-k; 98.80.Bp}

\maketitle
\section{Introduction}
Since its development in the beginning of the $20$th century, General Relativity has been the most successful theory
to describe the gravitational interaction. Nevertheless, several problems arise when one tries to construct models
in agreement with observational data in the classical approach. Among such problems we can mention, for
instance, the cosmological constant problem\cite{wein}, the cosmic coincidence\cite{papa} and the dark
sector\cite{mukhanov} - whose existence has been until today inferred only via its gravitational effects.
While these are still open issues in cosmology nowadays,
the prediction of an initial singularity for our Universe\cite{wald} is a crucial pathology from General Relativity.
\par Notwithstanding the cosmic censorship conjecture\cite{penrose}, there is no doubt that General Relativity must be
properly corrected or even replaced by a completely new theory. This demand
is necessary to solve the issue of the presence of the initial singularity predicted by classical General
Relativity. In this context, the initial conditions from which our Universe has evolved should depend crucially on the
adopted version of the theory to describe the dynamics around the singularity.
\par One of the most important features of our Universe as inferred by observational data is the
high degree of homogeneity and isotropy on large scales. However, when we consider a homogeneous and
isotropic model filled with baryonic matter, we find several difficulties by taking into account the primordial state of
our Universe. Among such difficulties, we can mention the horizon and flatness problems \cite{mukhanov}. Although the
Inflationary Paradigm\cite{abbott} allows to solve problems like these, inflationary cosmology does not solve the
initial singularity problem.
\par During the last decades, bouncing models\cite{novello} have been considered in order to circumvent the problem of
the initial singularity predicted by General Relativity. Such models (as \cite{nelson1}-\cite{mns}) might provide attractive
alternatives to the inflationary paradigm once they are able to solve the horizon and flatness problems, and
justify the power spectrum of primordial cosmological perturbations inferred by observations.
\par In the framework of bouncing cosmologies, it has been shown\cite{wands1,wands2} that curvature perturbations
develop an almost scale invariant spectrum if the contracting phase is dominated by a dust-like fluid and the
perturbations are generated by quantum vacuum fluctuations. However, in \cite{wands2} it was shown that the Bardeen
potential \cite{bardeen} grows larger than 1 at the bounce and this inconsistent behaviour can be avoided by a suitable
gauge choice.
\par In this paper we consider a Brane World scenario\cite{maeda} where a non-compact timelike extra dimension is introduced
\cite{sahni}. At low energies General Relativity is recovered but at high energy scales significant changes
are introduced into the gravitational dynamics and the singularities can be removed\cite{mns,sahni,maier}.

\par
While spacelike extra dimensions theories have received
more attention in the last decades\cite{maartens}, studies
involving extra timelike dimensions have been
considered\cite{sak} despite the fact that propagating
tachyonic modes or negative norm states may arise
due to timelike extra dimensions. These modes have
been regarded as problematic once they might violate
causality\cite{dvali} by considering interactions among usual
particles. Issues like the exceedingly small lower bound
on the size of timelike extra dimensions\cite{ynd}, the imaginary
self-energy of charged fermions induced by tachyonic
modes – which seems to cause disappearance of fermions
into nothing – and the spontaneous decay of stable particles
induced by tachyons with negative energy are major
difficulties\cite{dvali}. Nevertheless, in order to address the cosmological
constant problem in Kaluza-Klein theories\cite{ya1}
or reconcile a solution of the hierarchy problem with the
cosmological expansion of the universe\cite{chai}, timelike extra
dimensions have been considered. On the other hand,
it has been shown in \cite{ya2} that the appearance of massless
ghosts in an effective four-dimensional theory can
be avoided by considering topological criteria in Kaluza-
Klein theories with extra compactified time-like dimensions.
Moreover, the avoidance of propagating tachyonic or
negative norm states can also be achieved by considering
a non-compact timelike extra dimension\cite{igl}, which is the
case in the models of this paper.

\par
In section~\ref{sect:FE} we present a brief review of the modified Einstein field equations in the
Brane World scenario. In section~\ref{sect:model}, we construct non-singular cosmological models with a
cosmological constant where the matter content on the brane is given by dust and radiation (cf.\cite{maier}). The phase space
structure of the background present two basic configurations, namely,
one bounce solutions or eternal universes. In section~\ref{sect:ScalPert},
we derive two dynamical systems for the scalar perturbations by considering two different assumptions -- a 5-D de Sitter bulk or a
5-D perturbed bulk.
In sections V we examine the evolution of scalar perturbations from a contracting phase through an expanding phase
in the case of one bounce orbits. We discuss and compare the growth of the scalar perturbations in
the two bulk configurations adopted (5-D de Sitter and 5-D perturbed) and examine its stability and bounded
behaviour in the bounce and in the late accelerated phase of one bounce solutions.
In section VI we examine the evolution of scalar perturbations in the case of eternal universes,
and how unstable these configurations are considering both assumptions about the bulk.
Final remarks and future perspectives are presented in the conclusions.
\section{Field Equations}\label{sect:FE}
\begin{figure}
\begin{center}
\includegraphics*[height=5cm,width=8.5cm]{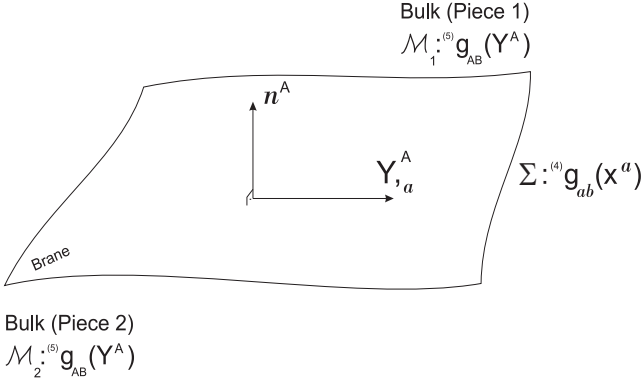}
\caption{The 4-D Lorentzian brane $\Sigma$ with metric
${^{(4)}g}_{ab}$, embedded in a 5-D conformally flat bulk $\cal{M}$ with metric ${^{(5)}g}_{AB}$. Capital Latin indices run from 0
to 4, small Latin indices run from 0 to 3. We regard $\Sigma$ as a common boundary of two pieces
${\cal{M}}_{1}$ and ${\cal{M}}_{2}$ of the bulk $\cal{M}$. While $Y^{A}(x^a)$ are the embedding functions of $\Sigma$ in
$\cal{M}$, the unit vector $n^{A}$ is normal to the boundary $\Sigma$ and has norm $\epsilon$.}
\label{fig:brane}
\end{center}
\end{figure}

For sake of completeness we give a brief introduction to Brane World theory, making explicit the specific assumptions
used to obtain the dynamics of the model. We refer to \cite{sahni} for a more complete and detailed discussion and
our notation closely follows \cite{wald}. We start with a 4-D Lorentzian brane $\Sigma$ with metric
${^{(4)}g}_{ab}$, embedded in a 5-D conformally flat bulk $\cal{M}$ with metric ${^{(5)}g}_{AB}$. Capital Latin indices run from 0
to 4, small Latin indices run from 0 to 3. We regard $\Sigma$ as a common boundary of two pieces
${\cal{M}}_{1}$ and ${\cal{M}}_{2}$ of $\cal{M}$ (cf. Fig.~\ref{fig:brane}) and the metric ${^{(4)}g}_{ab}$ induced on the brane
by the metric of the two
pieces should coincide although the extrinsic curvatures of $\Sigma$ in ${\cal{M}}_{1}$ and ${\cal{M}}_{2}$ can be
different. The action for the theory has the general form

\begin{equation}\label{eqn:eq4r}
\begin{split}
S=\frac{1}{2\kappa^2_{5}}\left\{\int_{M_{1}}\sqrt{-\epsilon~^{(5)}g}\left[^{(5)}R-2\Lambda_{5}
+2\kappa^2_{5}L_{5}\right]d^5x\phantom{^{(5)}g}\right.\\
+\int_{M_{2}}\sqrt{-\epsilon~^{(5)}g}\left[^{(5)}R-2\Lambda_{5}+2\kappa^2_{5}L_{5}\right]d^5x\phantom{^{(5)}g}\\
\left.\phantom{^{(5)}g}+2\epsilon\int_{\Sigma}\sqrt{-^{(4)}g}K_{2}d^4x-2\epsilon\int_{\Sigma}\sqrt{-^{(4)}g}K_{1}
d^4x\right\}\phantom{^{(5)}g}\\
+\frac{1}{2}\int_{\Sigma}\sqrt{-^{(4)}g}\left(\frac{1}{2\kappa^2_{4}}^{(4)}R-2\sigma\right)d^4x\phantom{^{(5)}g}\\
+\int_{\Sigma}\sqrt{-^{(4)}g}L_{4}({^{(4)}g}_{ab},\rho)d^4x\;.\phantom{^{(5)}g}
\end{split}
\end{equation}
In the previous equation, $^{(5)}R$ is the Ricci scalar of the Lorentzian 5-D metric $g_{AB}$ on $\cal{M}$, and
$^{(4)}R$ is the scalar curvature of the induced metric ${^{(4)}g}_{ab}$ on $\Sigma$. The parameter $\sigma$ denotes the
brane tension. The unit vector $n^{A}$ is normal to the boundary $\Sigma$ and has norm $\epsilon$. If
$\epsilon=+1$ the signature of the bulk space is $(+,+,-,-,-)$, so that the extra dimension is timelike. The quantity
$K=K_{ab}{^{(4)}g}^{ab}$ is the trace of the symmetric tensor of extrinsic curvature
$K_{ab}=Y_{,a}^{C}~Y_{,b}^{D}~\nabla_{C}n_{D}$, where $Y^{A}(x^a)$ are the embedding functions of $\Sigma$ in
$\cal{M}$\cite{eisenhart}.
While $L_{4}({^{(4)}g}_{ab},\rho)$ represents the Lagrangian density of the perfect fluid\cite{taub} (with equation of state
$p=\alpha\rho$), whose dynamics is restricted to the brane $\Sigma$, $L_{5}$ denotes the Lagrangian of matter in the
bulk. All integrations over the bulk and the brane are taken with the natural volume elements
$\sqrt{-\epsilon~{^{(5)}}g}~d^{5}x$ and $\sqrt{-{^{(4)}}g}~d^{4}x$ respectively. Einstein constants in five- and
four-dimensions are indicated with $\kappa^2_{5}$ and $\kappa^2_{4}$, respectively. Throughout the paper we use
natural units with $\hbar=c=1$.

\par Variations that leave the induced metric on $\Sigma$ intact, furnish the equations

\begin{equation}\label{eqn:eq6r}
^{(5)}G_{AB}+\Lambda_5~{^{(5)}}g_{AB}=\kappa^2_5~{^{(5)}}T_{AB}\;.
\end{equation}
Considering arbitrary variations of ${^{(5)}g}_{AB}$ and taking into account Eq.~(\ref{eqn:eq6r}), we obtain

\begin{equation}\label{eqn:eq7r}
^{(4)}G_{ab}+\epsilon\frac{\kappa^2_4}{\kappa^2_5}\left(S^{(1)}_{ab}+S^{(2)}_{ab}\right)
=\kappa^2_4\left(\tau_{ab}-\sigma g_{ab}\right)\;,
\end{equation}
where $S_{ab}\equiv K_{ab}-K{^{(4)}g}_{ab}$. In the limit
$\kappa^2_4\gg\kappa^2_5$,
Eq.~(\ref{eqn:eq7r}) reduces to
the Israel-Darmois junction condition{\bf s}\cite{israel}

\begin{equation}\label{eqn:eq9r}
\left(S^{(1)}_{ab}+S^{(2)}_{ab}\right)=-\epsilon\kappa^2_5\left(\tau_{ab}-\sigma {^{(4)}g}_{ab}\right)\;.
\end{equation}

Imposing the $Z_2$-symmetry\cite{maartens} and using the junction conditions (Eq.~\ref{eqn:eq9r}), we determine the
extrinsic curvature on the brane,
\begin{equation}\label{eqn:eq10r}
K_{ab}=\frac{\epsilon}{2} \kappa^2_5 \left[\left(\tau_{ab}-\frac{1}{3}\tau {^{(4)}g}_{ab}\right)+\frac{\sigma}{3}
{^{(4)}g}_{ab}\right]\;.
\end{equation}

\par Now using Gauss equation

\begin{equation}\label{eqn:gauss}
\begin{split}
^{(4)}R_{abcd}=^{(5)}R_{MNRS} Y^{M}_{,a} Y^{N}_{,b} Y^{R}_{,c} Y^{S}_{,d}-\\
\epsilon \left(K_{ac}K_{bd}-K_{ad}K_{bc} \right)\;,
\end{split}
\end{equation}
together with Eqs.~(\ref{eqn:eq6r}) and (\ref{eqn:eq10r}) we obtain the induced field equations on the brane

\begin{equation}\label{eqn:eq1.2.13}
\begin{split}
^{(4)}G_{ab}+\Lambda_{4}{^{(4)}g}_{ab}=8\pi G_{N}\tau_{ab}-\epsilon\kappa^4_{5}\Pi_{ab}+\\
\epsilon{E}_{ab}+\epsilon F_{ab}\;.
\end{split}
\end{equation}
In the above $E_{ab}={^{(5)}}C_{ABCD}n^A Y^{B}_{,a}n^{C} Y^{D}_{,b}$ is the projection of the 5-D Weyl tensor, and we have defined

\begin{eqnarray}\label{eqn:eq1.2.14}
\Lambda_{4}
& = & \frac{1}{2}\kappa^2_{5}\left(\frac{\Lambda_{5}}{\kappa^2_5}-\frac{1}{6}\epsilon\kappa^2_{5}\sigma^2\right)\;,\\
G_{N} & = & \epsilon\frac{\kappa^4_{5}\sigma}{48\pi}\;,\\
\nonumber
\Pi_{ab} & = & -\frac{1}{4}\tau_{a}^{c}\tau_{bc} +\frac{1}{12}\tau\tau_{ab}
+\frac{1}{8}{^{(4)}g}_{ab}\tau^{cd}\tau_{cd}-\\
& & \frac{1}{24}\tau^2{^{(4)}g}_{ab}\;,\\
\nonumber
F_{ab} & = & \frac{2}{3}\kappa^2_{5}\Big\{\epsilon ~{^{(5)}T}_{BD} Y^B_{,a} Y^D_{,b}-\\
& & \Big[{^{(5)}T}_{BD} n^{B} n^{D}+\frac{1}{4}\epsilon~{^{(5)}T}\Big]{^{(4)}g}_{ab}\Big\}\;,
\end{eqnarray}
where $G_N$ is simply the Newton's constant on the brane.
Here we stress that the effective 4-D cosmological constant can be set to zero
in the present case of an extra timelike dimension by properly fixing the bulk cosmological constant as
$\Lambda_{5}=\frac{1}{6}\kappa_{5}^{4} ~\sigma^2$.
It is important to notice that for a 4-D brane embedded in a conformally flat bulk we have the absence of the Weyl
conformal tensor projection $E_{ab}$, and $F_{ab}$ in Eq.~(\ref{eqn:eq1.2.13}).

On the other hand, Codazzi's equations imply that
\begin{equation}\label{eqn:eq1.2.17}
\nabla_{a} K-\nabla_{b}K^{b}_{a}=\frac{1}{2}\epsilon \kappa^2_{5}\nabla_{b}\tau^{b}_{a}\;.
\end{equation}
By imposing that $\nabla_{b}\tau^{b}_{a}=0$, the Codazzi conditions read
\begin{equation}\label{eqn:eq1.2.18}
\nabla^{a}{E}_{ab}=\kappa^4_{5}\nabla^{a}\Pi_{ab}+\nabla^{a}F_{ab}.
\end{equation}
where $\nabla_a$ is the covariant derivative with respect to the induced metric ${^{(4)}g}_{ab}$. Eqs.~(\ref{eqn:eq1.2.13}) and
(\ref{eqn:eq1.2.18}) are the dynamical equations of the gravitational field on the brane.

\section{The Model}\label{sect:model}

We now consider a closed FLRW geometry on the four-dimensional brane embedded in a five-dimensional conformally
flat bulk with a timelike extra dimension.
As explicitly shown in \cite{sahni}, the conformally flat bulk is described by a 5D de Sitter spacetime
with metric
\begin{eqnarray}
\label{eqn:eq1.2.181}
\nonumber
ds^2=
\Big(k-\frac{\Lambda_5}{6} v^2\Big)du^2+\frac{1}{\Big(k-\frac{\Lambda_5}{6} v^2\Big)}dv^2
\\
-v^2\Big[\frac{1}{1-kr^2}dr^2+r^2(d\theta^2+\sin^2{\theta}d\varphi^2)\Big],
\end{eqnarray}
where $(r,\theta,\varphi)$ are the comoving coordinates and $k$ is the $3$-curvature of the spatial sections.
The equation $v=a(u)$ describes the position of the brane.

\par
The matter content of the model is given by dust and radiation which are
confined to the brane. A complete treatment of the phase space dynamics of these models, including
their non-integrable extensions when a scalar field is introduced, was given in \cite{maier}. Considering comoving
coordinates on the brane, the line element is given by

\begin{equation}\label{eqn:eq2}
ds^2=a^2(\eta)\left\{d\eta^2-\left[\frac{1}{1-kr^2}dr^2+r^2d\Omega^2\right]\right\},
\end{equation}
where $\eta$ is the conformal time, $a(\eta)$ the scale factor and $d\Omega^2= d \theta^2+ \sin^2 \theta d \varphi^2$.

\par
\begin{figure}
\begin{center}
\includegraphics*[height=5.5cm,width=8.5cm]{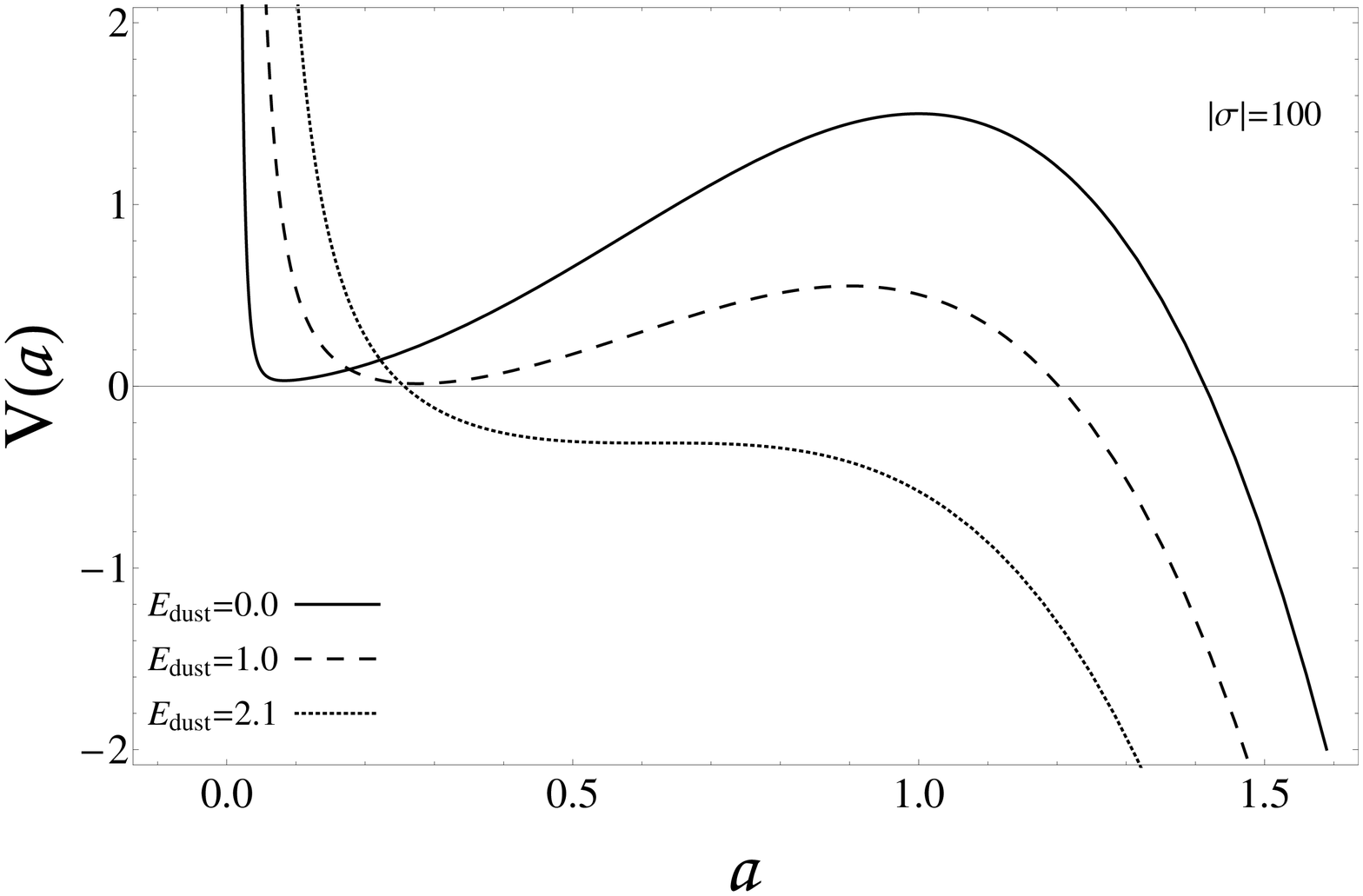}
\caption{The potential $V(a)$ for $\Lambda=1.5$, $E_{\rm rad}=0.01$, $|\sigma|=100$ and several values of $E_{\rm
dust}$. For higher values of $E_{\rm dust}$, namely $E_{\rm dust} \gtrsim 2.1$, the minimum of the potential is no
longer present and the two extremal of the potential vanish.}
\label{fig:potential}
\end{center}
\end{figure}

 From (\ref{eqn:eq1.2.13}) we obtain the modified Friedmann equations,

\begin{equation}\label{eqn:eq4}
\begin{split}
{\cal H}^2+k-\frac{\Lambda_{4}a^2}{3} & = \frac{\kappa^2_4a^2}{3}(\rho_{\rm dust}+\rho_{\rm rad})\\
& \times\left[1-\frac{1}{{2|\sigma|}}(\rho_{\rm dust}+\rho_{\rm rad})\right]\;,
\end{split}
\end{equation}
where ${\cal H}=\frac{{a}^{\prime}}{a}$ with $a^{\prime}=\frac{da}{d\eta}$, and $\kappa^2_4=8\pi G_N$. We define the
energy momentum tensor as $\tau^{ab}$=$\tau^{ab}_{\rm dust}+\tau^{ab}_{\rm rad}$, assuming that
$\nabla_{b}\tau^{ab}_{\rm dust}=0=\nabla_{b}\tau^{ab}_{\rm rad}$. Therefore, for the non-interacting perfect fluids
above, we have

\begin{equation}\label{eqn:eq3}
\rho_{\rm dust}=\frac{E_{\rm dust}}{a^3}\;,\quad\rho_{\rm rad}=\frac{E_{\rm rad}}{a^4}
\end{equation}
where $E_{\rm dust}$ and $E_{\rm rad}$ are both constants of motion.
By defining
\begin{equation}\label{eqn:eq3.5}
\begin{split}
V(a)=3ka^2-\Lambda_{4}a^4-\kappa^2_4 a E_{\rm dust}+\\
\frac{\kappa^2_4}{2|\sigma|} \left(\frac{E_{\rm dust}}{a}+\frac{E_{\rm rad}}{a^2}\right)^2\;,
\end{split}
\end{equation}
the first integral of motion (Eq.~\ref{eqn:eq4}) is given by the Hamiltonian constraint
\begin{equation}\label{eqn:eq3.6}
\frac{p^2_a}{12}+V(a)=\kappa^2_4 E_{\rm rad}\;,
\end{equation}
where $p_a\equiv 6a^{\prime}$. Here it is easy to see (cf. Fig.~\ref{fig:potential}) that although $V(a)\rightarrow -\infty$ when
$a\rightarrow
+\infty$, the last terms in the potential (Eq.~\ref{eqn:eq3.5}) act as an infinite potential barrier which
avoids the singularity at $a=0$.
\par From equation Eq.~(\ref{eqn:eq3.6}) we derive the following dynamical system equivalent to the modified Einstein
equations on the brane:

\begin{equation}\label{eqn:eq5}
a^{\prime}=\frac{p_a}{6}\;,\quad p^{\prime}_a=-\frac{dV}{da}\;.
\end{equation}

\par The critical points in the phase-space are stationary solutions of Eq.~(\ref{eqn:eq5}), namely, the points
$(a=a_{\rm crit},p_a=0)$, at which the RHS of Eqs.~(\ref{eqn:eq5}) vanishes. Here, $a_{\rm crit}$ are the real
positive roots of $dV/da$.
In the remaining of the
paper we will restrict ourselves to spatially closed models
$k>0$ and fix $k=1$.
By considering the case of closed geometries ($k>0$), it is not difficult to verify that,
depending on the values of the parameters $(\Lambda_4,|\sigma|,E_{\rm rad},E_{\rm dust})$, there are at most two
critical points\cite{maier} in the phase-space connected with one local minimum and one local maximum of $V(a)$. In this
case the minimum of the potential corresponds to a centre $P_0$ while the maximum corresponds to a saddle point $P_1$
(see Fig~\ref{fig:potential}).
\begin{figure}[t]
\begin{center}
\includegraphics*[height=5.5cm,width=8.5cm]{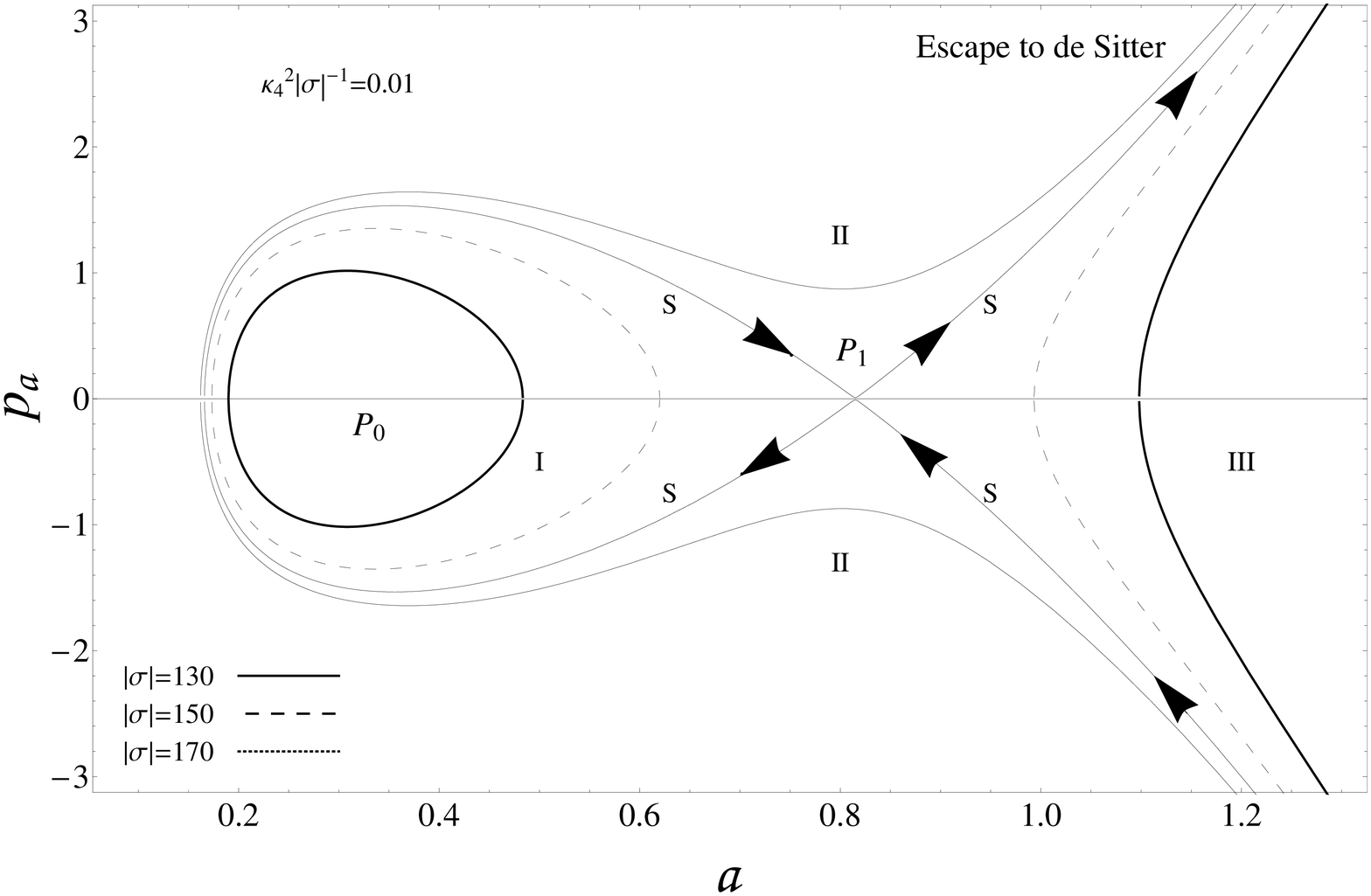}
\caption{The phase-space for $\Lambda=1.5$, $E_{\rm rad}=0.01$ and $E_{\rm dust}=1.0$. The critical points $P_0$
(centre) and $P_1$ (saddle) correspond to stable and unstable Einstein universes respectively. The periodic orbits of
region I describe perpetually bouncing universes. Orbits in Region II and III are solutions of one bounce universes. A
separatrix $\mathcal{S}$ emerges from $P_1$ towards the de Sitter attractor defining an escape to de Sitter.}
\label{fig:phasespace}
\end{center}
\end{figure}
This configuration allows us to obtain different types of orbits that describe the evolution of different
universes in this model. Typically, this model allows for the presence of eternal universes (region I)
and one bounce universes (regions II and III) as shown in Fig.~\ref{fig:phasespace}.
In the present paper we will focus our attention to eternal universes and the one bounce
solutions of regions I and II, respectively. The one bounce solutions of region III, although
mathematically sound, do not have an immediate interest since they cannot provide
a concrete cosmological bouncing model connected with the observations as orbits of region II can.
Actually orbits of region II exhibit two accelerated phases (the first following immediately the bounce and the
second corresponding to a late de Sitter acceleration phase) separated by a smooth transition corresponding to a decelerated
expansion phase and compatible with observational parameters, as shown in \cite{mns}.

\section{Scalar Perturbations}\label{sect:ScalPert}

Let us consider scalar perturbations on the brane in the longitudinal gauge\cite{mukhanov}

\begin{equation}\label{eqn:eqsp1}
ds^2=a^2(\eta)[(1+2\phi)d\eta^2-(1-2\psi)\gamma_{ij}dx^idx^j]\;,
\end{equation}
where
\begin{equation}\label{eqn:eqsp2}
\gamma_{ij}\equiv {\rm diag}\left[\frac{1}{1-r^2}dr^2,r^2,r^2\sin^2\theta\right]\;.
\end{equation}

In order to derive the hydrodynamical perturbations, we assume that the energy momentum tensor in the background
(Eq.~\ref{eqn:eq2}) is given by

\begin{equation}\label{eqn:eqsp01}
\tau^{ab}\equiv\tau^{ab}_{\rm dust}+\tau^{ab}_{\rm rad}\;,
\end{equation}
where
\begin{equation}\label{eqn:eqsp02}
\tau^{a}_{\phantom{a}b~\rm dust}= \rho_{\rm dust}V^{a}V_{b}\;,
\end{equation}
and
\begin{equation}\label{eqn:eqsp03}
\tau^{a}_{\phantom{a}b~\rm rad}= \frac{4}{3}\rho_{\rm rad}V^{a}V_{b}-
\frac{1}{3}\rho_{\rm rad}\delta^{a}_{\phantom{a}b}\;,
\end{equation}
and $V_a\equiv a\delta^{0}_{a}$ is the background velocity field. By considering scalar hydrodynamical perturbations,
we have that

\begin{equation}\label{eqn:eqsp04}
\delta\tau^{a}_{\phantom{a}b~\rm dust}=\delta\rho_{\rm dust}V^{a}V_{b}
+\delta q^{a}_{\rm dust}V_{b}+\delta q_{b~\rm dust}V^{a}\;,
\end{equation}
and
\begin{equation}\label{eqn:eqsp05}
\begin{split}
\delta\tau^{a}_{\phantom{a}b~\rm rad}=\frac{4}{3}\delta\rho_{\rm rad}V^{a}V_{b}-
\frac{1}{3}\delta\rho_{\rm rad}\delta^{a}_{\phantom{a}b}+\delta q^{a}_{\rm rad}V_{b}\\
+\delta q_{b~\rm rad}V^{a}+\delta \pi^{a}_{\phantom{a}b~\rm rad}\;.
\end{split}
\end{equation}

Here we are assuming that the equation of state $\delta p_N=w_N\delta\rho_N$ ($N=\rm dust, \rm rad$) holds and that the
scalar hydrodynamical perturbations of the dust component do not induce an anisotropic pressure. On the other hand we
have defined the perturbed heat flux as

\begin{equation}\label{eqn:eqsp05nn}
\delta q^{a}_{~N}=(0,-a\nabla^{a}(\delta q_{N}))=(0,-a(\rho_{0N}+p_{0N})(\nabla^{a}v_N))\;,
\end{equation}
where $\delta q_{N}$ and $v$ are scalars. Furthermore, the perturbed anisotropic pressure of radiation is defined as

\begin{equation}\label{eqn:eqsp05n}
\delta \pi^{i}_{\phantom{i}j~\rm rad}= {^{(3)}\nabla^{i}}{^{(3)}\nabla_{j}}\delta\pi_{\rm rad}
-\frac{1}{3a^2}\delta^{i}_{\phantom{i}j} \Delta\delta \pi_{\rm rad}\;,
\end{equation}
where $\delta\pi_{rad}$ is a scalar, ${^{(3)}\nabla}_{i}$ is the covariant derivative built with the metric
$\gamma_{ij}$ and $\Delta\equiv g^{ij} {^{(3)}\nabla}_{i} {^{(3)}\nabla}_{j}$.

\par Assuming that the $Z_2$ symmetry is not broken as we introduce scalar perturbations, we obtain,

\begin{equation}\label{eqn:eqnew}
\delta K^{a}_{\phantom{a}b}=\frac{\epsilon}{2} \kappa^2_5 \left(\delta\tau^{a}_{\phantom{a}b}-\frac{1}{3}\delta\tau
\delta^{a}_{\phantom{a}b}\right)\;.
\end{equation}

In this context, the perturbed Einstein field equations on the brane are given by

\begin{equation}\label{eqn:eq1.2.13.new}
\delta^{(4)}G^{a}_{\phantom{a}b}=8\pi G_{N}
\left(\delta\tau^{a}_{\phantom{a}b}-\frac{6\epsilon}{|\sigma|}\kappa^4_{5}\delta\Pi^{a}_{\phantom{a}b}\right)
+\epsilon\delta{E}^{a}_{\phantom{a}b}\;.
\end{equation}

\par In general, the perturbed spacetime (Eq.~\ref{eqn:eqsp1}) can be embedded in a non-conformally flat bulk so
perturbations of the Weyl tensor $\delta{E}^{\alpha}_{\phantom{\alpha}\beta}$ must correct the perturbed field equations
(cf. Eq.~\ref{eqn:eq1.2.13.new}). As $\delta{E}^{\alpha}_{\phantom{\alpha}\beta}$ is a symmetric traceless tensor, it mimics
a radiation component with heat flux and anisotropic pressure\cite{maartens}, that is,
\begin{equation}\label{eqn:eqsp06}
\begin{split}
\delta E^{a}_{\phantom{a}b}=\frac{4}{3}\delta\rho_{E}V^{a}V_{b}-\frac{1}{3}\delta\rho_{E}\delta^{a}_{~b}+\\
\delta q^{a}_{E}V_{b}+\delta q_{E~b}V^{a}+\delta \pi^{a}_{E~b}\;.
\end{split}
\end{equation}
Therefore the perturbed Einstein field equations read
\begin{equation}\label{eqn:eqsp8}
\begin{split}
\Delta\psi-3{\cal H}(\psi^{\prime}+{\cal H}\phi)+3\psi
=\frac{\kappa_4^2a^2}{2}\left\{(\delta\rho_{\rm dust}+\delta\rho_{\rm rad})\phantom{\frac{1}{|\sigma|}}\right.\\
\left.\times\left[1-\frac{1}{|\sigma|}(\rho_{\rm dust}+\rho_{\rm rad})\right]-\delta\rho_E\right\}\;,
\end{split}
\end{equation}
\begin{equation}\label{eqn:eqsp9}
\begin{split}
\psi^{\prime}+{\cal H}\phi
=-\frac{\kappa_4^2a^2}{2}\left\{(\delta q_{\rm dust}+\delta q_{\rm rad})\phantom{\frac{1}{|\sigma|}}\right.\\
\left.\times\left[1-\frac{1}{|\sigma|}(\rho_{\rm dust}+\rho_{\rm rad})\right]-\delta q_E\right\}\;,
\end{split}
\end{equation}
and
\begin{equation}\label{eqn:eqsp10}
\begin{split}
(2{\cal H}^{\prime}+{\cal H}^2)\phi+{\cal H}\phi^{\prime}+\psi^{\prime\prime}+2{\cal H}\psi^{\prime}-\psi+
\frac{1}{3}\Delta D~~~~~~~~~~~~~~~~~~~~~~~~~~\\
=\frac{\kappa_4^2a^2}{2}\left\{\frac{1}{3}\delta\rho_{\rm rad}-\frac{1}{|\sigma|}\left[(\delta \rho_{\rm dust}+
\delta\rho_{\rm rad})
\left(\rho_{\rm dust}+\frac{4}{3}\rho_{\rm rad}\right)\phantom{\frac{1}{3}}\right.\right.~~~~~~~~\\
+\left.\left.\frac{1}{3}\delta\rho_{\rm rad}(\rho_{\rm dust}+\rho_{\rm rad})\right]-\frac{1}{3}\delta\rho_E\right\}\;.
~~~~~~~~~~~~~~~~~~~~~~
\end{split}
\end{equation}
where $D\equiv \phi-\psi$.

\par On the other hand, from the conservation equations
$\nabla_{b}\delta\tau^{cb}_{\rm dust}=0=\nabla_{b}\delta\tau^{cb}_{\rm rad}$, we obtain

\begin{equation}\label{eqn:eqsp3}
\delta\rho^{\prime}_{\rm dust}=-3{\cal H}\delta\rho_{\rm dust}+3\psi^{\prime}\rho_{\rm dust}-
\Delta\delta q_{\rm dust}\;,
\end{equation}

\begin{equation}\label{eqn:eqsp30}
\delta\rho^{\prime}_{\rm rad}=-4{\cal H}\delta\rho_{\rm rad}+4\psi^{\prime}\rho_{\rm rad}-\Delta\delta q_{\rm rad}\;,
\end{equation}
for $c=0$, and
\begin{equation}\label{eqn:eqsp4}
\delta q^{\prime}_{\rm dust}=-4{\cal H}\delta q_{\rm dust}-\rho_{\rm dust}\phi\;,
\end{equation}

\begin{equation}\label{eqn:eqsp5}
\begin{split}
\delta q^{\prime}_{\rm rad}=-4{\cal H}\delta q_{\rm rad}-\frac{4}{3}\rho_{\rm rad}\phi-\frac{1}{3}\delta\rho_{\rm rad}\\
-\frac{2}{3a^2}(\Delta+3)\delta\pi_{\rm rad}\;,
\end{split}
\end{equation}
for $c=1,2,3$. Imposing that $\nabla_{b}\delta E^{0b}=0$ and $\nabla_{b}\delta E^{ib}=\kappa^4_5\nabla_{b}\delta\Pi^{ib}$,
we respectively obtain
\begin{equation}\label{eqn:eqsp6}
\delta\rho^{\prime}_{E}=-4{\cal H}\delta \rho_{E}-\Delta\delta q_{E}\;,
\end{equation}
and
\begin{equation}\label{eqn:eqsp7}
\begin{split}
\delta q^{\prime}_{E}+4{\cal H}\delta q_{E}+\frac{\delta\rho_E}{3}
+\frac{2}{3a^2}(\Delta+3)\delta\pi_{E}\\
=-\frac{1}{|\sigma|}
\left(\rho_{\rm dust}+\frac{4}{3}\rho_{\rm rad}\right)\\
\times\left[\delta\rho_{\rm dust}+\delta\rho_{\rm rad}-3{\cal H}(\delta q_{\rm dust}+
\delta q_{\rm rad})\phantom{\frac{1}{a^2}}\right.\\
-\left.\frac{1}{a^2}(\Delta+3)\delta\pi_{\rm rad}\right]\;.
\end{split}
\end{equation}

\par Finally we will expand the scalar perturbation variables in scalar eigenmodes of the Laplacian
operator $\Delta$ on $S^3$, $Q ^{N}_{lm}(r,\theta,\varphi)=R^{N}_{l}(r) Y_{lm}(\theta,\varphi)$,
which satisfy
\begin{eqnarray}\label{eqn:eqapa}
\Delta Q ^{N}_{lm}=(N^2-1)Q ^{N}_{lm}\;,
\end{eqnarray}
where $N$ is a positive integer. The multiplicity of each mode $N$ is $N^2$.
The two modes $N=1$ and $N=2$ are gauge modes since they respectively correspond to a change in the curvature radius
(homogeneous deformation) and in a displacement of the centre of the 3-sphere and are
thus physically not relevant\cite{weeks,lifshitz}.
Scalar perturbations can be examined
using its {\it Fourier} components, or modes, $({\delta {\mathcal T}})_{Nlm}(\eta)~ Q ^{(N)}_{lm}(r,\theta,\varphi)$, where
$\delta {\mathcal T}$ stands for $\psi$, $\delta \rho_{\rm dust}$, $\delta \rho_{\rm rad}$, and so on.
In this instance Eqs. (\ref{eqn:eqsp8})-(\ref{eqn:eqsp7}) reduce to ordinary differential equations
for the variables $\psi_{_{Nlm}}(\eta)$, ${\delta \rho_{\rm dust}}_ {_{Nlm}}(\eta)$, ${\delta \rho_{\rm rad}}_{_{Nlm}}(\eta)$,
etc., decoupled for each mode. Each $N$ defines a comoving scale for the modes.
We should note that for the case of a non-normalized 3-curvature $k$ the comoving scale of the modes
is $[k (N^2-1)]^{-1/2}$. Most of our numerical analysis in the paper will consider only the physical modes $N \geq 3$.
In the following sections the dynamics of the scalar perturbations will be examined via these Fourier variables.
We will drop the mode indices in order to avoid overcluttering in the equations and formulae.

\par From the embedding perspective, two basic assumptions can be made about the bulk in which
we consider the perturbed FLRW brane. We now examine these two possibilities in order to
derive a close dynamical system which can be numerically evolved.

\subsection{The case of a de Sitter Bulk}

In order to simplify our analysis, we will first examine the case where the spacetime (Eq.~\ref{eqn:eqsp1}) is
embedded in a de Sitter bulk. As we know, such assumption cannot be ruled out from a mathematical point of view.
In fact, by imposing $\delta\rho_{E}=\delta q_{E}=\delta\pi_{E}=0$, the embedding of the perturbed spacetime
Eq. (\ref{eqn:eqsp1}) in a de Sitter bulk is still valid as long as the Gauss-Codazzi equations
(\ref{eqn:eqsp8})-(\ref{eqn:eqsp7}) hold. In this case, Eq.~(\ref{eqn:eqsp7}) provides the simple constraint for
hydrodynamical perturbations

\begin{equation}\label{eqn:eqsp7new}
\delta\rho_{\rm dust}+\delta\rho_{\rm rad}-3{\cal H}(\delta q_{\rm dust}+\delta q_{\rm rad})
=\frac{1}{a^2}(N^2+2)\delta\pi_{\rm rad}\;.
\end{equation}

On the other hand, from Eqs. (\ref{eqn:eqsp8}) and Eq. (\ref{eqn:eqsp9}) we derive the further constraint
\begin{equation}\label{eqn:eqv2}
\begin{split}
(N^2+2)\psi
=\frac{\kappa_4^2a^2}{2}\left\{\left[1-\frac{1}{|\sigma|}(\rho_{\rm dust}+\rho_{\rm rad})\right]\right.\\
\left.\phantom{\frac{1}{|\sigma|}}\times\left[\delta\rho_{\rm dust}+\delta\rho_{\rm rad}-3{\cal H}(\delta q_{\rm dust}+
\delta q_{\rm rad})\right]\right\}\;.
\end{split}
\end{equation}
Now, from Eqs. (\ref{eqn:eqsp7new}) and (\ref{eqn:eqv2}) we obtain
\begin{eqnarray}\label{eqn:eq1n}
{\kappa^2_4}\delta\pi_{\rm rad}=\frac{2\psi|\sigma|}{|\sigma|-\rho_{\rm dust}-\rho_{\rm rad}}\;.
\end{eqnarray}
We remark that by taking the derivative of Eq.~(\ref{eqn:eqsp9}) and substituting it in Eq.~(\ref{eqn:eqsp10}), it results
\begin{equation}\label{eqn:eqspn}
(\psi-\phi)={\kappa^2_4}\delta\pi_{\rm rad}\left[1+\frac{1}{2|\sigma|}(\rho_{\rm dust}+2\rho_{\rm rad})\right]\;,
\end{equation}
where we have used Eqs.~(\ref{eqn:eqsp4}), (\ref{eqn:eqsp5}) and (\ref{eqn:eqsp7new}). Therefore, inserting
Eq.~(\ref{eqn:eq1n}) into Eq.~(\ref{eqn:eqspn}), one can easily show that the relation
\begin{eqnarray}\label{eqn:eq2n}
\phi=\left(\frac{2\rho_{\rm dust}+3\rho_{\rm rad}+|\sigma|}{\rho_{\rm dust}+\rho_{\rm rad}-|\sigma|}\right)\psi\;,
\end{eqnarray}
holds, fixing the dynamics of $\phi$ in terms of the dynamics of $\psi$ or vice-versa.
\par From a simple examination of the dynamical system (\ref{eqn:eqsp9})-(\ref{eqn:eqsp7}) we can see that
the dynamics of the variable $\delta \pi_{\rm rad}$ is not determined. However the use of Eq. (\ref{eqn:eq1n})
allows us to eliminate the variable $\delta \pi_{\rm rad}$ from the dynamical system.
It is worth noticing that we could also eliminate the variables $\delta \pi_{\rm rad}$ by
assuming that the radiation anisotropic pressure is proportional to the shear according to the following relation
\begin{eqnarray}\label{eqn:eqsh}
\delta\pi_{\alpha\beta~\rm rad}&=&\mu\delta\sigma_{\alpha\beta~\rm rad}\nonumber\\
&=&\mu\delta\left(\nabla_{(\beta}V_{\alpha)}-\dot{V}_{(\alpha}V_{\beta)}-\frac{1}{3}\theta h_{\alpha\beta}\right)\;,
\end{eqnarray}
where $\mu$ is a constant and

\begin{eqnarray}\label{eqn:eqsh1}
\nabla_{(\beta}V_{\alpha)}&\equiv&\frac{1}{2}(\nabla_{\beta}V_{\alpha}+\nabla_{\alpha}V_{\beta})\\
\dot{V}_{\alpha}V_{\beta}&\equiv&\frac{1}{2}[(\nabla_{\gamma}{V}_{\alpha})V^{\gamma}V_{\beta}+
(\nabla_{\beta}{V}_{\gamma})V^{\gamma}V_{\alpha}]\\
\theta&\equiv&\nabla_{\alpha}V^{\alpha}\\
h_{\alpha\beta}&\equiv& g_{\alpha\beta}-V_{\alpha}V_{\beta}\;.
\end{eqnarray}

In this case $\delta \sigma_{00~\rm rad}=\delta \sigma_{0i~\rm rad}=0$. On the other hand, from
Eqs.~(\ref{eqn:eqsp05nn}) and (\ref{eqn:eqsp05n})

\begin{equation}\label{eqn:eqsh3}
\delta\sigma_{ij~\rm rad}=-\frac{3a}{4\rho_{\rm rad}}\left[{^{(3)}\nabla}_{i}{^{(3)}\nabla}_{j}\delta
q_{\rm rad}-\frac{1}{3}\gamma_{ij}\Delta\delta q_{\rm rad}\right]\;.
\end{equation}

Thus, according to Eq.~(\ref{eqn:eqsh}):

\begin{equation}\label{eqn:eqsh4}
\delta\pi_{\rm rad}=-\frac{3\mu a}{4\rho_{\rm rad}}\delta q_{\rm rad}\;,
\end{equation}
so that a dynamical system with six degrees of freedom $(a, p_a, \phi, \delta\rho_{\rm dust}, \delta q_{\rm dust},
\delta\rho_{\rm rad})$ would be obtained. However, in order to perform a more general analysis, we will make no such
assumption. In this case we end up with the following dynamical system
\begin{eqnarray}
\label{eqn:dsds}
a^\prime & = & \frac{p_a}{6}\;,\\
p^\prime_a & = & -6a+4\Lambda_4a^3+\kappa^2_4E_{\rm dust}\nonumber\\
& & +\frac{\kappa^2_4}{|\sigma|}\left(\frac{E_{\rm dust}}{a}+
\frac{E_{\rm rad}}{a^2}\right)\left(\frac{E_{\rm dust}}{a^2}+\frac{2E_{\rm rad}}{a^3}\right),~~
\end{eqnarray}

\begin{eqnarray}\label{eqn:dsds}
\psi^{\prime} & = & -\frac{p_a}{6a}\phi\nonumber\\
& & -\frac{\kappa_4^2a^2}{2}(\delta q_{\rm dust}+\delta q_{\rm rad})
\left[1-\frac{1}{|\sigma|}(\rho_{\rm dust}+\rho_{\rm rad})\right]\;,\\
\delta\rho^{\prime}_{\rm dust} & = & -\frac{p_a}{2a}(\delta \rho_{\rm dust}+\phi\rho_{\rm dust})\nonumber\\
& & -\frac{3\kappa_4^2a^2}{2}(\delta q_{\rm dust}+\delta q_{\rm rad})
\left[1-\frac{1}{|\sigma|}(\rho_{\rm dust}+\rho_{\rm rad})\right]\rho_{\rm dust}\nonumber\\
& & -(N^2-1)\delta q_{\rm dust}\;,\\
\delta\rho^{\prime}_{\rm rad} & = & -\frac{2p_a}{3a}(\delta \rho_{\rm rad}+\phi\rho_{\rm rad})\nonumber\\
& & -\kappa_4^2a^2(\delta q_{\rm dust}+\delta q_{\rm rad})
\left[1-\frac{1}{|\sigma|}(\rho_{\rm dust}+\rho_{\rm rad})\right]\rho_{\rm rad}\nonumber\\
& & -(N^2-1)\delta q_{\rm rad}\;,\\
\delta q^{\prime}_{\rm dust} & = & -\frac{2p_a}{3a}\delta q_{\rm dust}-\rho_{\rm dust}\phi\;,\\
\delta q^{\prime}_{\rm rad} & = & -\frac{2p_a}{3a}\delta q_{\rm rad}-\frac{4}{3}\rho_{\rm rad}\phi-\frac{1}{3}\delta
\rho_{\rm rad}\nonumber\\
& & -\frac{2}{3a^2}(N^2+2)\delta\pi_{\rm rad}\;,
\end{eqnarray}
where we have used Eqs.~(\ref{eqn:eq3.5}), (\ref{eqn:eq5}), (\ref{eqn:eqsp9})-(\ref{eqn:eqsp5}). It is worth remarking again
that $\phi=\phi(\psi,a)$ and $\pi_{\rm rad}=\pi_{\rm rad}(\psi,a)$ according to Eqs.~(\ref{eqn:eq1n}) and
(\ref{eqn:eq2n}), respectively, so that a dynamical system with seven degrees of freedom $(a, p_a, \psi, \delta\rho_{\rm dust}, \
delta q_{\rm dust},
\delta\rho_{\rm rad}, \delta q_{\rm rad})$ is obtained. Furthermore, the constraint equations
(\ref{eqn:eq3.6}) and (\ref{eqn:eqv2}) must be satisfied during the evolution of the above dynamical system.
In fact these two constraints are used to check the accuracy of our numerical evaluations in the next sections.
In all our evaluations the error in the constraints were always $\leq 10^{-10}$.

\subsection{The Case of a Perturbed Bulk}

To proceed we will now examine the consequences of assuming a perturbed bulk. In this case, the
$5$-D scalar perturbations would effectively correct the $4$-dimensional perturbed equations
with a traceless Weyl fluid. As we saw in Eq. (\ref{eqn:eqsp06}), we can decompose this fluid in its energy density
$\delta \rho_E$, heat flux $\delta q_E$ and anisotropic pressure $\delta \pi_E$. Taking the derivative of equation
Eq.~(\ref{eqn:eqsp9}) and substituting in Eq.~(\ref{eqn:eqsp10}), we obtain

\begin{equation}\label{eqn:pb1}
(\psi-\phi)={\kappa^2_4}\left\{\delta\pi_{\rm
rad}\left[1+\frac{1}{2|\sigma|}(\rho_{\rm dust}+2\rho_{\rm rad})\right]-\delta\pi_{E}\right\}\;,
\end{equation}
where we have used Eqs.~(\ref{eqn:eqsp4}), (\ref{eqn:eqsp5}) and (\ref{eqn:eqsp7}). As there is no dynamical equation
for the anisotropic pressures $\delta\pi_{\rm rad}$ and $\delta\pi_{E}$, the dynamical system to be
evolved is not a closed system unless $\delta\pi_{\rm rad}=\delta\pi_{E}=0$. In this case, $\phi=\psi$
and we obtain the following dynamical system
\begin{eqnarray}\label{pb2}
a^\prime & = & \frac{p_a}{6}\;,\\
p^\prime_a & = & -6a+4\Lambda_4a^3+\kappa^2_4E_{\rm dust}\nonumber\\
& & +\frac{\kappa^2_4}{|\sigma|}\left(\frac{E_{\rm dust}}{a}+
\frac{E_{\rm rad}}{a^2}\right)\left(\frac{E_{\rm dust}}{a^2}+\frac{2E_{\rm rad}}{a^3}\right)\;,\\
\phi^{\prime} & = & -\frac{p_a}{6a}\phi-\frac{\kappa_4^2a^2}{2}\left\{(\delta q_{\rm dust}+
\delta q_{\rm rad})\phantom{\frac{1}{|\sigma|}}\right.\nonumber\\
& & \left.\times\left[1-\frac{1}{|\sigma|}(\rho_{\rm dust}+\rho_{\rm rad})\right]-\delta q_E\right\}\;~~
\end{eqnarray}
\begin{eqnarray}\label{pb2}
\delta\rho^{\prime}_{\rm dust} & = & -\frac{p_a}{2a}(\delta\rho_{\rm dust}+\phi\rho_{\rm dust})\nonumber\\
& & -\frac{3\kappa_4^2a^2}{2}\left\{(\delta q_{\rm dust}+
\delta q_{\rm rad})\phantom{\frac{1}{|\sigma|}}\right.\nonumber\\
& & \left.\times\left[1-\frac{1}{|\sigma|}(\rho_{\rm dust}+\rho_{\rm rad})\right]-
\delta q_E\right\}\rho_{\rm dust}\nonumber\\
& & -(N^2-1)\delta q_{\rm dust}\;,\\
\delta\rho^{\prime}_{\rm rad} & = & -\frac{2p_a}{3a}(\delta\rho_{\rm rad}+\phi\rho_{\rm rad})\nonumber\\
& & -\kappa_4^2a^2\left\{(\delta q_{\rm dust}+\delta q_{\rm rad})\phantom{\frac{1}{|\sigma|}}\right.\nonumber\\
& & \left.\times\left[1-\frac{1}{|\sigma|}(\rho_{\rm dust}+\rho_{\rm rad})\right]-
\delta q_E\right\}\rho_{\rm rad}\nonumber\\
& & -(N^2-1)\delta q_{\rm rad},~~~~
\end{eqnarray}
\begin{eqnarray}
\delta\rho^{\prime}_{E} & = & -\frac{2p_a}{3a}\delta\rho_{E}-(N^2-1)\delta q_{E}\;,\\
\delta q^{\prime}_{\rm dust} & = & -\frac{2p_a}{3a}\delta q_{\rm dust}-\rho_{\rm dust}\phi\;,\\
\delta q^{\prime}_{\rm rad} & = & -\frac{2p_a}{3a}\delta q_{\rm rad}-\frac{4}{3}\rho_{\rm rad}\phi-
\frac{1}{3}\delta\rho_{\rm rad}\;,\\
\delta q^{\prime}_{E} & = & -\frac{2p_a}{3a}\delta q_{E}-\frac{\delta\rho_E}{3}-\frac{1}{|\sigma|}
\left(\rho_{\rm dust}+\frac{4}{3}\rho_{\rm rad}\right)\nonumber\\
& & \times\left[\delta\rho_{\rm dust}+\delta\rho_{\rm rad}-\frac{p_a}{2a}(\delta q_{\rm dust}+
\delta q_{\rm rad})\right]\;.
\end{eqnarray}
where we have used Eqs.~(\ref{eqn:eq3.5}), (\ref{eqn:eq5}) and (\ref{eqn:eqsp9})-(\ref{eqn:eqsp7}). Together
with Eq.~(\ref{eqn:eq3.6}), the constraint for scalar perturbations
\begin{equation}\label{eqn:pb3}
\begin{split}
(N^2+2)\phi=\frac{\kappa_4^2a^2}{2}\left\{\left[1-\frac{1}{|\sigma|}(\rho_{\rm dust}+\rho_{\rm rad})\right]\right.\\
\times\left[\delta\rho_{\rm dust}+\delta\rho_{\rm rad}-\frac{p_a}{2a}(\delta q_{\rm dust}+\delta q_{\rm rad})\right]\\
\left.\phantom{\frac{1}{|\sigma|}}-\delta\rho_E+\frac{p_a}{2a}\delta q_E\right\}\;,
\end{split}
\end{equation}
obtained from substituting Eq.~(\ref{eqn:eqsp9}) in Eq.~(\ref{eqn:eqsp8}), must also be satisfied during the evolution
of the above dynamical system. We will also use this constraint together with the background constraint (\ref{eqn:eq3.6})
to check the validity of our numerical evaluations of the dynamics in the present model.
In all our evaluations the error in the constraints were always, again, $\leq 10^{-10}$.

\section{One Bounce Solutions}\label{sect:dSBulk}

In the context of bouncing cosmologies an important question which arises is whether the bounce spoils or not scalar
perturbations. The one-bouncing solutions of region III, although mathematically sound, will
not be considered here since they
lack an important feature of the one bounce orbits of region II, that is the presence of a
smooth transition phase of decelerated expansion connecting two acceleration phases.
Actually it has been shown in \cite{mns} that one of the orbits like those in region II of Fig.~\ref{fig:phasespace}
provide a concrete cosmological bouncing model, namely, given observational parameters the structure of the
phase space in region II shows a nonsingular orbit with
two accelerated phases, separated by a smooth transition corresponding to a decelerated expansion.
Such phases can be connected to a primordial accelerated phase, a soft transition to Friedmann (where the classical regime is valid),
and a graceful exit to a de Sitter accelerated phase. Motivated by this observational feature, in
this section we will restrict ourselves to one bounce orbits in region II only.

\par In this first approach, we intend to solve numerically the equations in order to understand how the bounce affects
the growth of scalar perturbations considering the transition from a contracting to an expanding phase.
As the dynamical systems obtained above  -- for a de Sitter or a perturbed bulk -- are extremely involved, the implementation of
larger values of $N$
from a numerical point of view is a task which demands
more powerful simulations which will be examined in the future. Nevertheless, from the dynamical point of view the qualitative
behaviour of the dynamics should be the same. Thus, as a first illustration we will restrict ourselves to large scale
perturbations (mostly $N=3, 4$ and $5$) in this section.

\subsection{The Evolution of Scalar Perturbations in a FLRW Brane Embedded in a de Sitter Bulk}

\par
\begin{figure}
\includegraphics*[height=5cm,width=8cm]{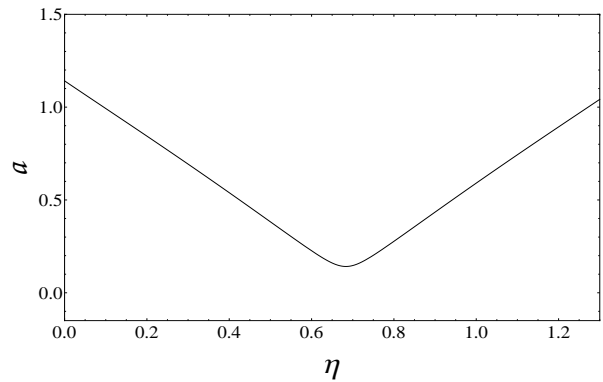}
\caption{The scale factor as a function of conformal time $\eta$. Here we fix the parameters $\kappa^2_4=|\sigma|=100,
~E_{\rm dust}=0.001,~E_{\rm rad}=0.08,~\Lambda=1.5$
and the initial condition $p_{a_0}=-9$.
The $16$ digit
precision values for the scale factor initial condition is given by $a_0=1.141525275590078$
so that the constraint (\ref{eqn:eq4}) is satisfied.
The scale of the bounce is $\eta_b\simeq 0.68$ and $a(\eta_b)\simeq 0.15$.}
\label{fig:a_eta}
\end{figure}

 Let us consider one given orbit in region II of Fig.~\ref{fig:phasespace}.
It can be shown that one of those orbits is generated by fixing the parameters and initial
conditions given in Fig.~\ref{fig:a_eta}.
In this case the scale of the bounce is $\eta_b\simeq 0.68$ so that $a(\eta_b)\simeq 0.15$ (see
Fig.~\ref{fig:a_eta}).

\begin{figure}
\includegraphics*[height=5cm,width=8.3cm]{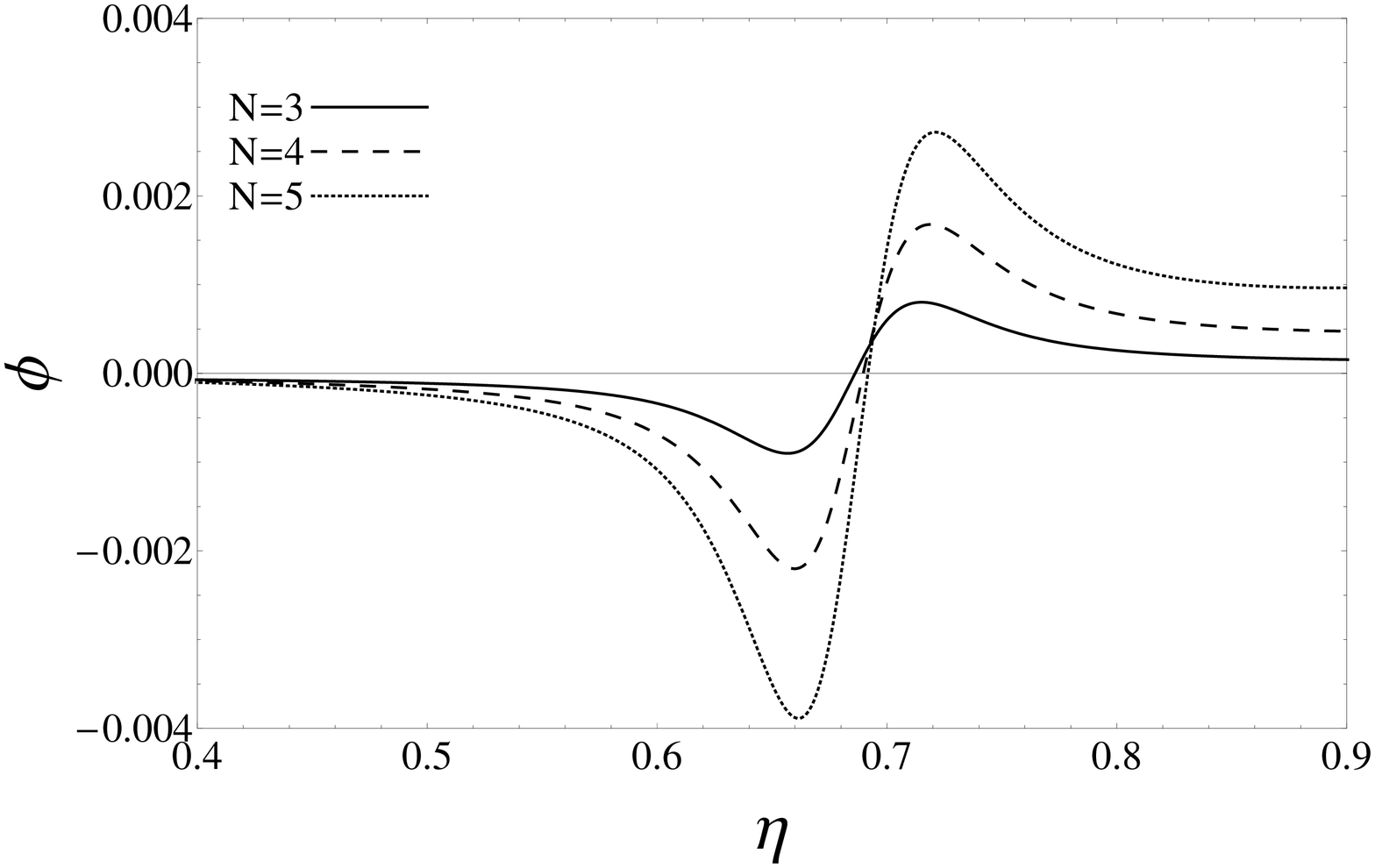}
\includegraphics*[height=5cm,width=8.5cm]{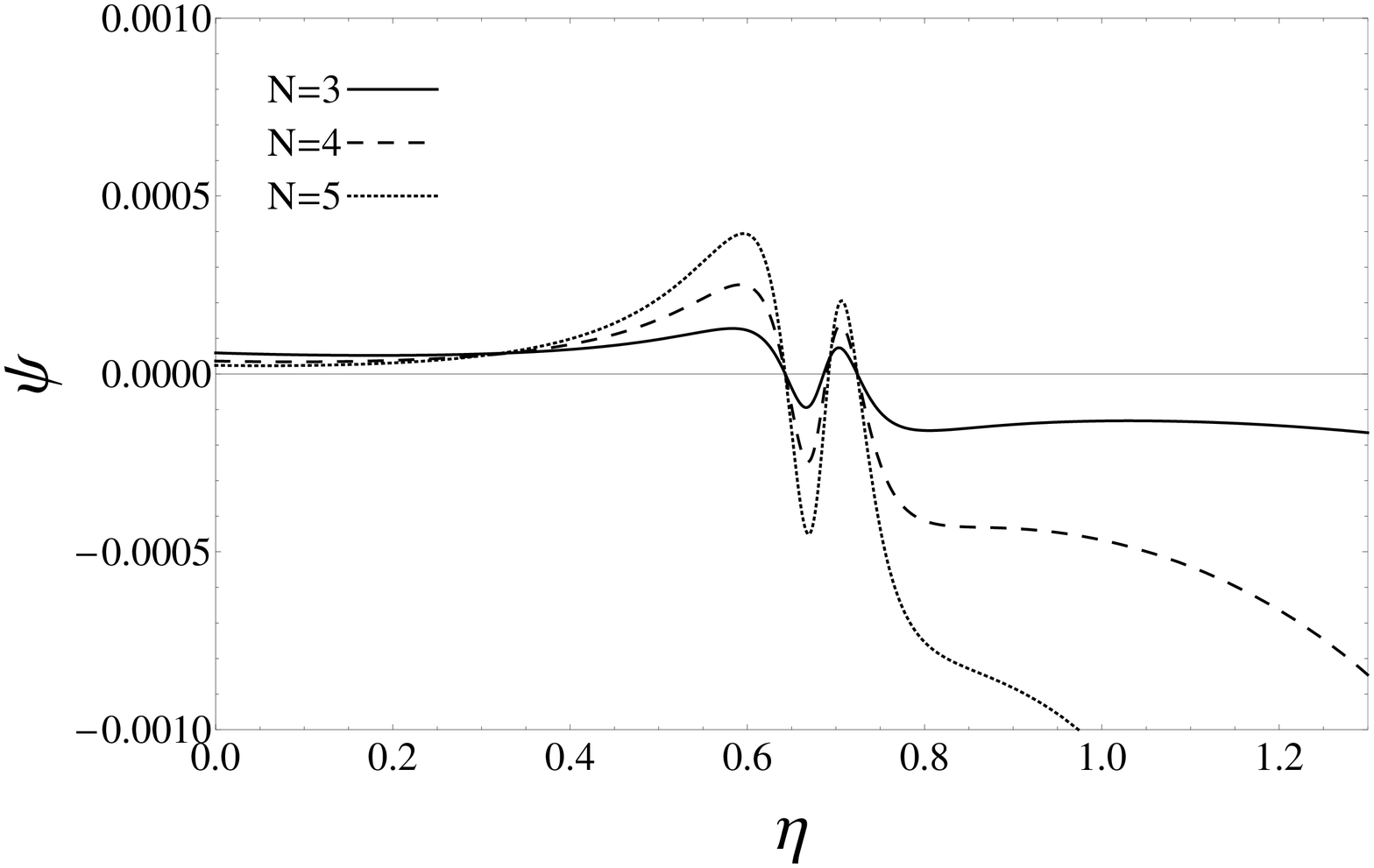}
\caption{The potentials $\phi$ (upper panel) and $\psi$ (lower panel) for $N=3,4,5$ (solid, dashed and dotted
curves, respectively). For the scalar perturbations, we adopt the initial conditions $\delta\rho_{\rm dust}(0)=\delta q_{\rm dust}
(0)=\delta q_{\rm rad}(0)=0$ and $\delta\rho_{\rm rad}(0)=10^{-5}$. The $16$ digit
precision values for the initial condition of $\psi$ is given by $\psi_0=0.000651228632253/(N^2+2)$
so that the constraint (\ref{eqn:eqv2}) is satisfied. In this case $|\phi| \ll 10^{-3}\ll 1$ and $|\psi| \simeq 10^{-4} \ll 1$ in a
finite neighbourhood $\Delta a $ of the bounce.}
\label{fig:PhiPsi}
\end{figure}

\begin{figure}
\includegraphics*[height=5cm,width=8.5cm]{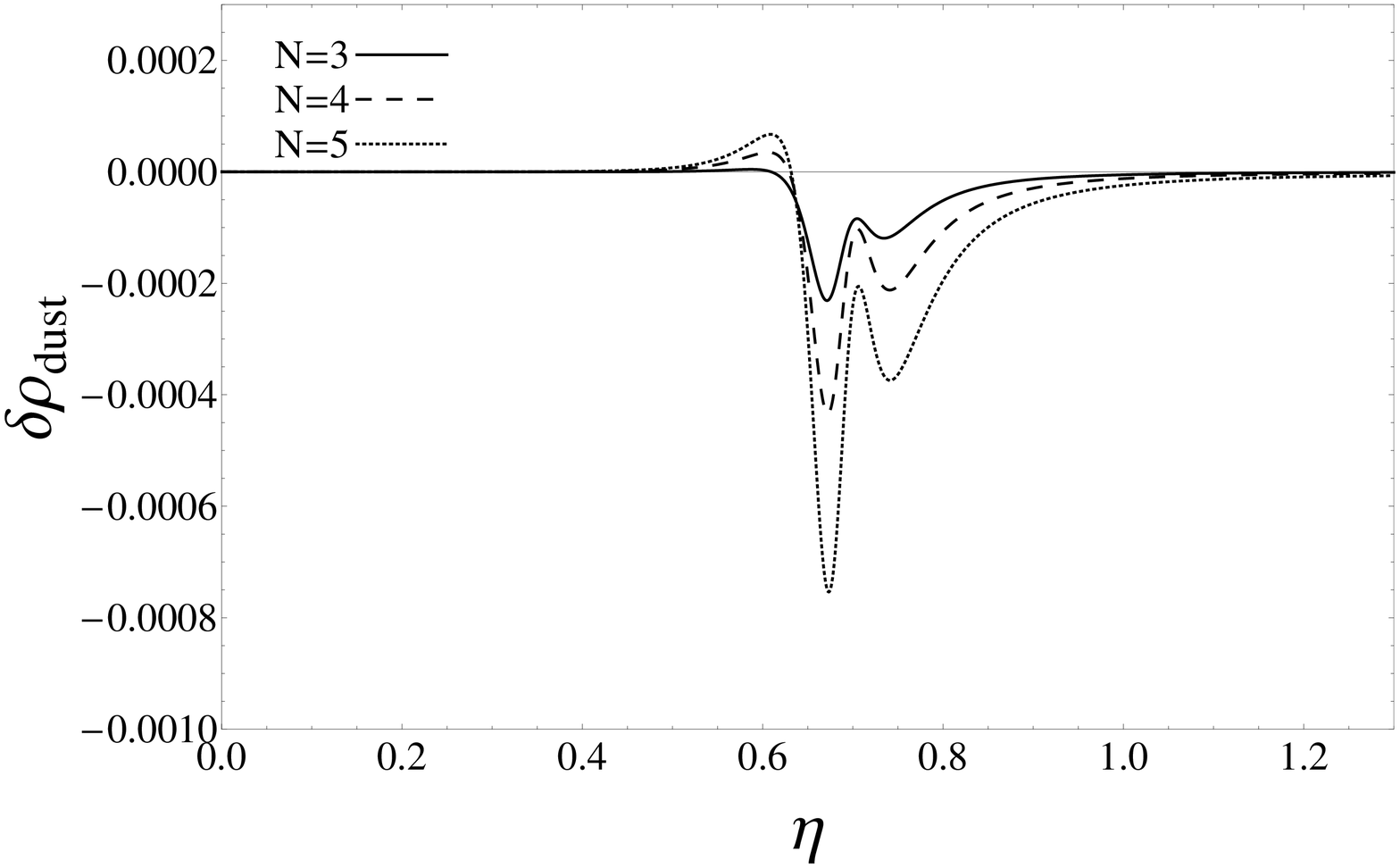}
\includegraphics*[height=5cm,width=7.9cm]{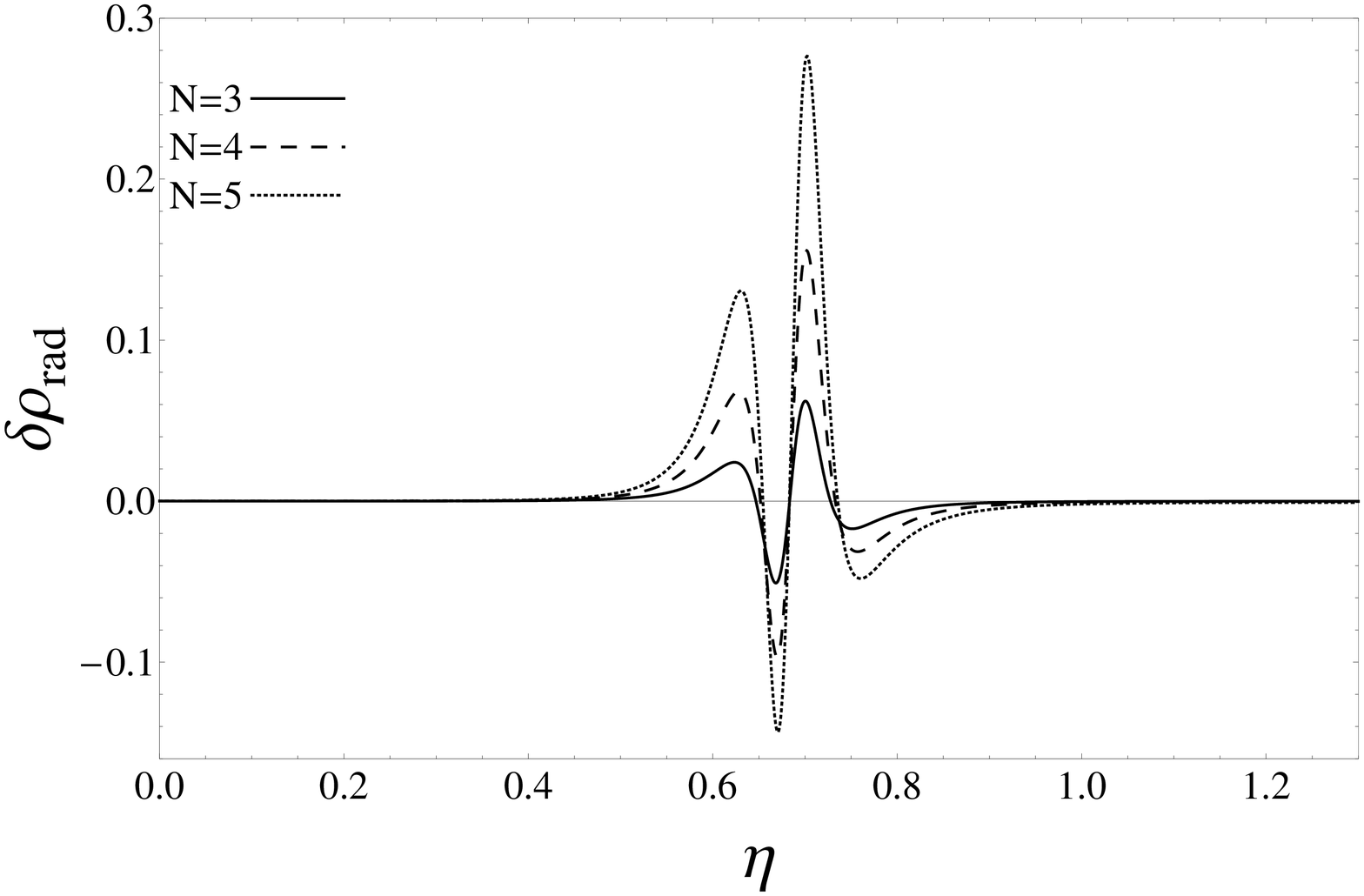}
\caption{Evolution of the energy density fluctuations $\delta\rho_{\rm dust}$ (upper panel) and $\delta\rho_{\rm
rad}$ (lower panel) for $N=3,4,5$ (solid, dashed and dotted, respectively). In this case the amplification of
$|\delta\rho_{\rm dust}|\leq 10^{-4}$ and $|\delta\rho_{\rm rad} (\eta_b)|\leq 0.1$.}
\label{fig:deltarho}
\end{figure}

\begin{figure}
\includegraphics*[height=5cm,width=8.2cm]{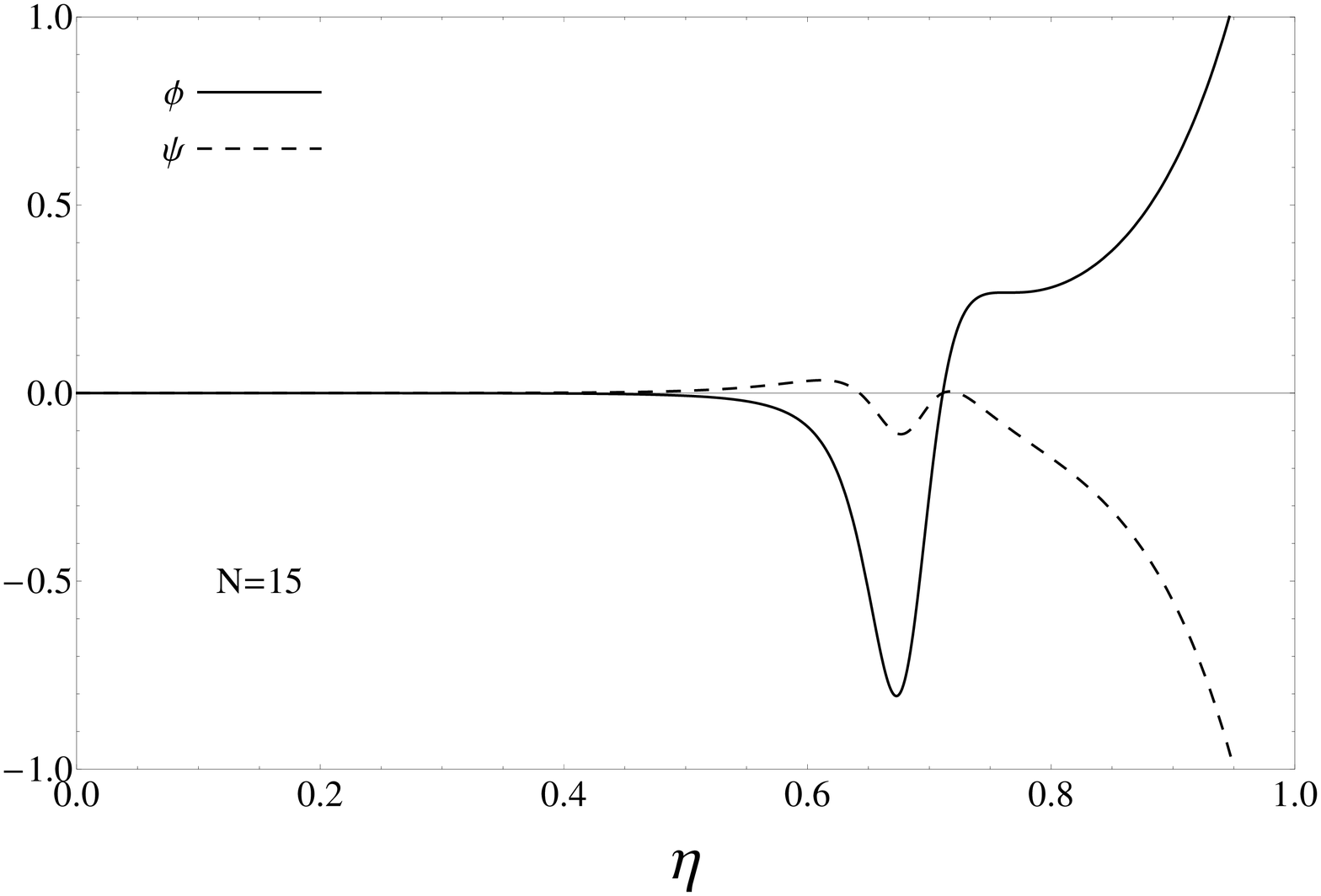}
\caption{The potentials $\phi$ and $\psi$ for $N=15$ (solid and dashed
curves, respectively). In this case $|\phi| \simeq 1$ in a
finite neighbourhood $\Delta a $ of the bounce, breaking the linear perturbative regime.}
\label{fig:new}
\end{figure}

\begin{figure}
\includegraphics*[height=5cm,width=8.2cm]{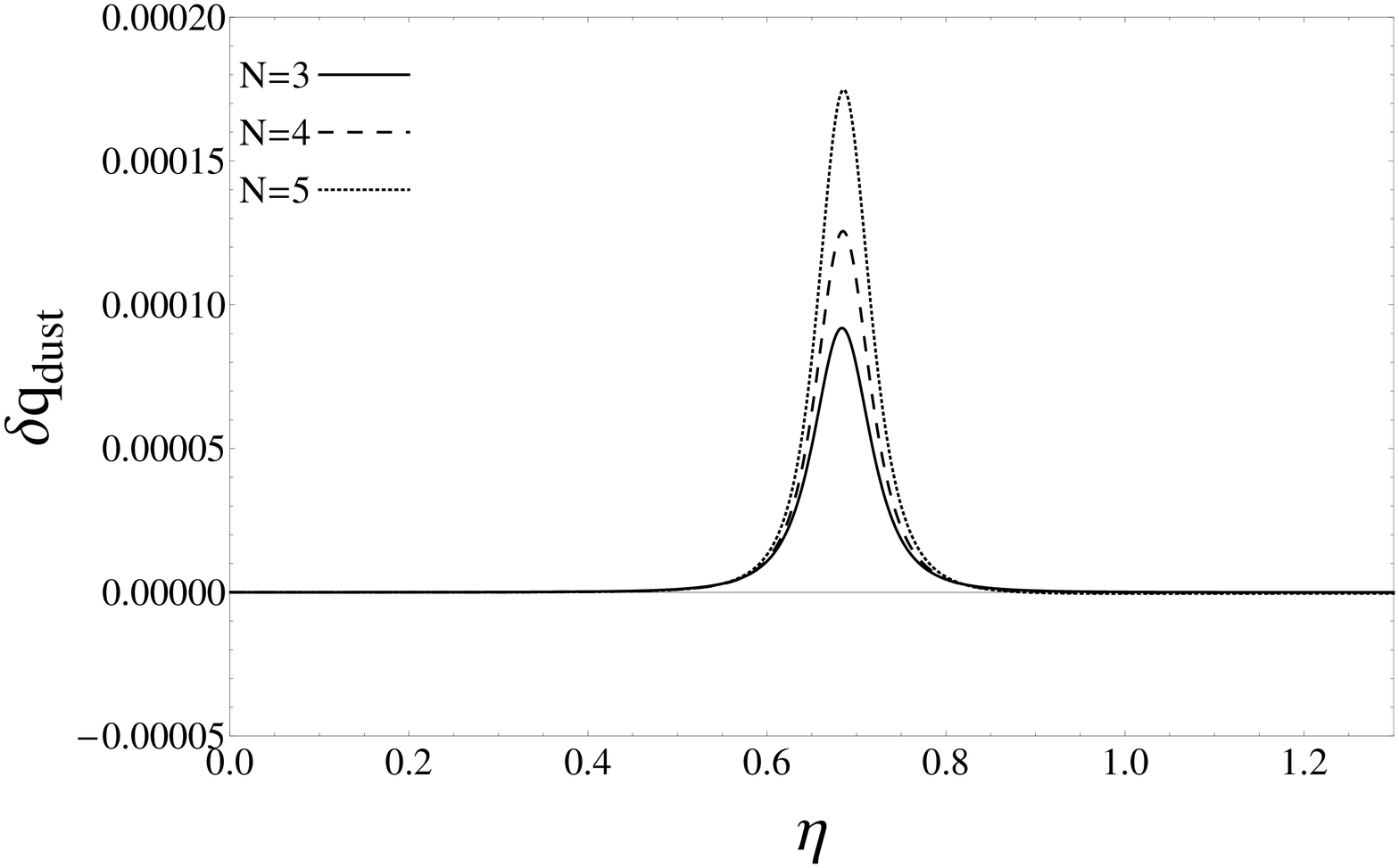}
\includegraphics*[height=5cm,width=8cm]{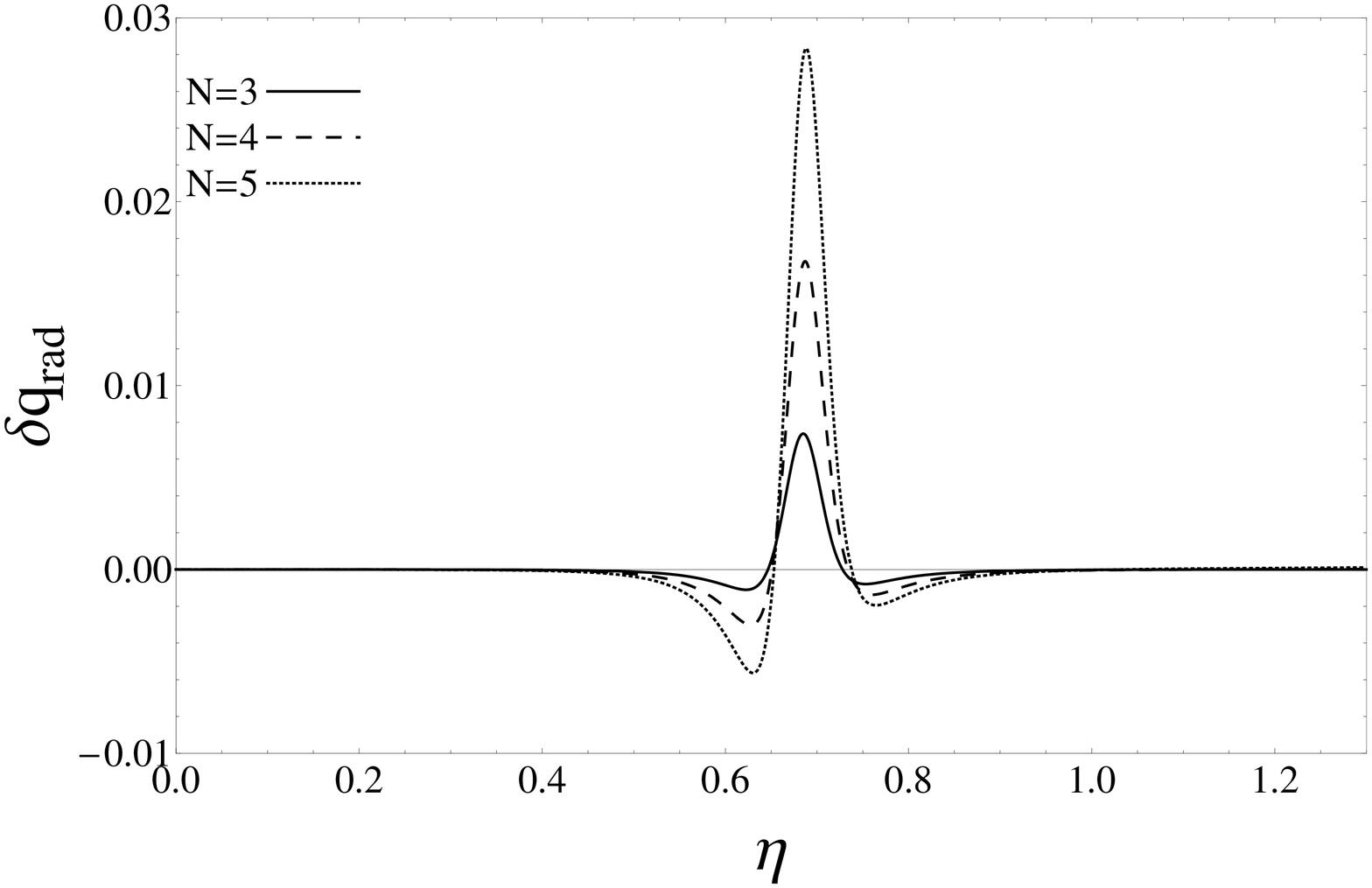}
\caption{Evolution of the heat flux perturbations: $\delta q_{\rm dust}$ (upper panel) and $\delta q_{\rm
rad}$ (lower panel) for $N=3,4,5$ (solid, dashed and dotted curves, respectively).}
\label{fig:deltaq}
\end{figure}

\begin{figure}
\includegraphics*[height=5cm,width=8.5cm]{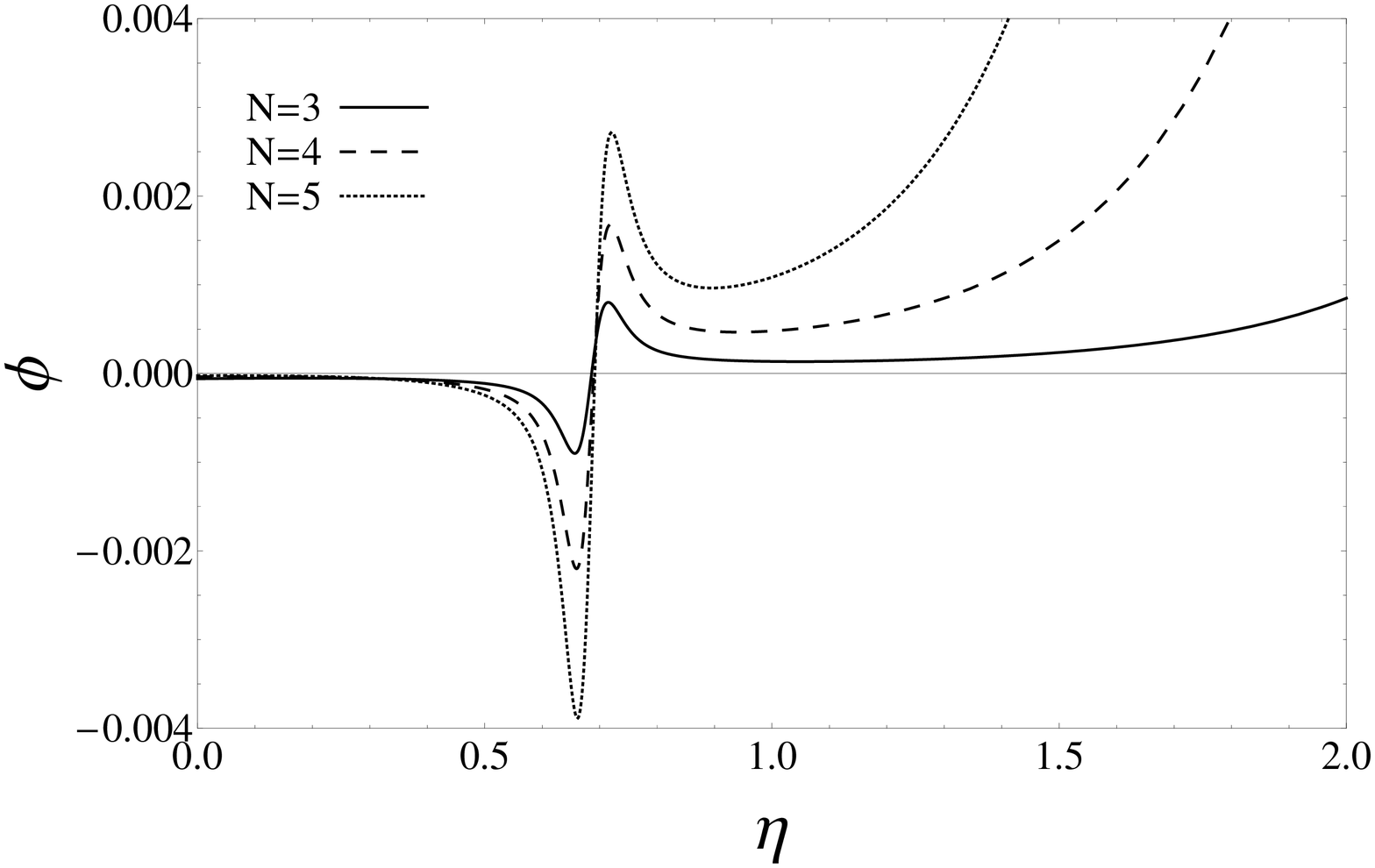}
\caption{Evolution of the potential $\phi$ in the case of a de Sitter bulk for $N=3,4,5$ (solid, dashed and dotted curves,
respectively). Here we see that perturbation variables increase with $N$ as
the universe enters the late accelerated phase ($\eta \simeq 1.2$), diverging as they approach the final de Sitter stage.}
\label{fig:phinew}
\end{figure}
\par In Fig.~\ref{fig:PhiPsi} we show the behaviour of the potentials $\phi$ (upper panel) and $\psi$ (lower
panel) for $N=3,4,5$ adopting a particular choice of initial conditions for the hydrodynamical perturbations. In this case $|\phi(
\eta)| \simeq 10^{-3}\ll 1$ and $|\psi(\eta)| \simeq 10^{-4} \ll 1$
in a finite neighbourhood of the bounce, namely, for any $a$ in the interval $\Delta a=a(\eta)-a(\eta_b) \lesssim 0.35 $.
Therefore the large-scale amplifications in the potentials are much less than $1$. In this sense the bounce does not
spoil the scalar metric perturbations connected to hydrodynamical fluctuations.

\par In Fig.~\ref{fig:deltarho} we show the behaviour of the energy density fluctuations for
$N=3,4,5$. In the upper (lower) panel we present results for $\delta\rho_{\rm dust}$ ($\delta\rho_{\rm rad}$).
In this case $|\delta\rho_{\rm rad}|\lesssim 0.1$ and $|\delta\rho_{\rm dust}| \lesssim 10^{-4}$ in the neighbourhood
$\Delta a$ of the bounce. Therefore we have

\begin{eqnarray}\label{eqn:eqpin2n}
\left|\frac{\delta \rho_{\rm rad}}{\rho_{\rm rad}}\right|_{\Delta a}\simeq \left|\frac{\delta
\rho_{\rm dust}}{\rho_{\rm dust}}\right|_{\Delta a} \simeq 10^{-4}\;.
\end{eqnarray}

We see that large-scale amplifications in dust and radiation energy densities are much smaller than their background values.
In this sense the bounce does not spoil the hydrodynamical perturbations connected to the energy densities.
However, this result hold only up to a certain scale. In fact, for $N=15$
we see from Fig.~\ref{fig:new} that $|\phi(\eta)| \simeq 1$ in a finite neighbourhood of the bounce, breaking the linear
perturbative regime.
\par Finally, for illustration purposes, we show in Fig.~\ref{fig:deltaq} the evolution of the heat flux perturbations, $\delta q_{\rm dust}$ in the
upper panel and $\delta q_{\rm rad}$ in the lower panel.
\par We should comment that, after the bounded amplification in the neighbourhood of the bounce, the scalar perturbations
maintain their bounded amplitudes in the expanding decelerated phase of the FLRW background after the bounce, until
the universe enters the late accelerated expansion phase towards the de Sitter configuration. In this late accelerated phase
an important feature of the scalar perturbations is their divergent growth, for any mode $N$, as the universe
approaches the de Sitter configuration (cf. Fig.~\ref{fig:phinew}).

\subsection{The Evolution of Scalar Perturbations in a FLRW Brane Embedded in a Perturbed Bulk}\label{sect:PBulk}

In order to examine how the assumption of a perturbed bulk modifies the growth of scalar perturbations
for one bounce universes (of region II cf. Fig.~\ref{fig:phasespace}), let us adopt for the background universe the same
parameters and initial
conditions assumed in the previous subsection.
As we have mentioned before, the scale of the bounce in this case is $\eta_b\simeq 0.68$ so that $a(\eta_b)\simeq
0.15$ (see Fig.~\ref{fig:a_eta}).

\par
\begin{figure}
\includegraphics*[height=5cm,width=8.5cm]{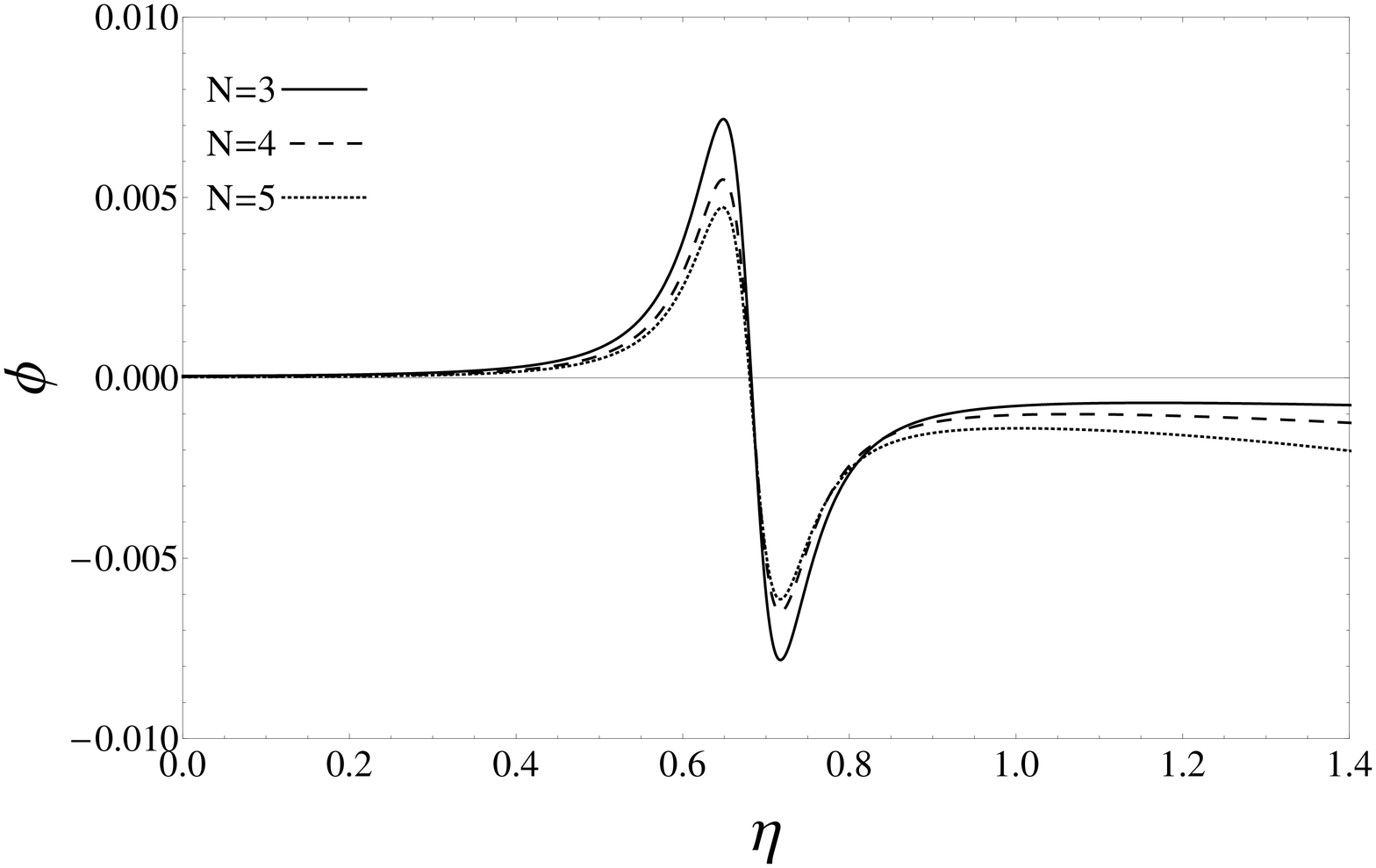}
\caption{The potential $\phi$ in the case of a perturbed bulk for $N=3,4,5$ (solid, dashed and dotted curves,
respectively). Here we fix
$\delta \rho_{\rm dust}(0)=\delta q_{\rm dust}(0)=\delta q_{\rm rad}(0)=\delta q_{E}(0)=0$, $\delta\rho_{\rm rad}(0)=10^{-5}$, $\
delta \rho_{E}(0)=10^{-6}$
as initial conditions for the scalar and hydrodynamical perturbations.
The $16$ digit
precision values for the initial condition of $\phi$ is given by $\phi_0=0.000586074634513/(N^2+2)$
so that the constraint (\ref{eqn:pb3}) is satisfied.
In this case $|\phi(\eta)| \leq 10^{-2}$ in any finite neighbourhood of the bounce.}
\label{fig:PhiPB}
\end{figure}

 In Fig.~\ref{fig:PhiPB} we show the behaviour of the potential $\phi$ for $N=3,4,5$
for a particular choice of initial conditions for the hydrodynamical perturbations.
We can see that in this case $|\phi(\eta)|
\lesssim 10^{-2}$ for $N=3,4,5$ in a finite neighbourhood of the bounce $\Delta a=|a(\eta)-a({\eta}_b)|$.
Fig.~\ref{fig:PhiPB} shows that the higher is the value of $N$ the smaller is the amplitude growth enhanced by the bounce.
Comparing with the upper panel of Fig.~\ref{fig:PhiPsi} we see that the amplification of $\phi$ by the bounce, in the
present case of a perturbed bulk, is about two orders of magnitude larger than that of a de Sitter bulk.

\begin{figure}
\includegraphics*[height=5cm,width=8.5cm]{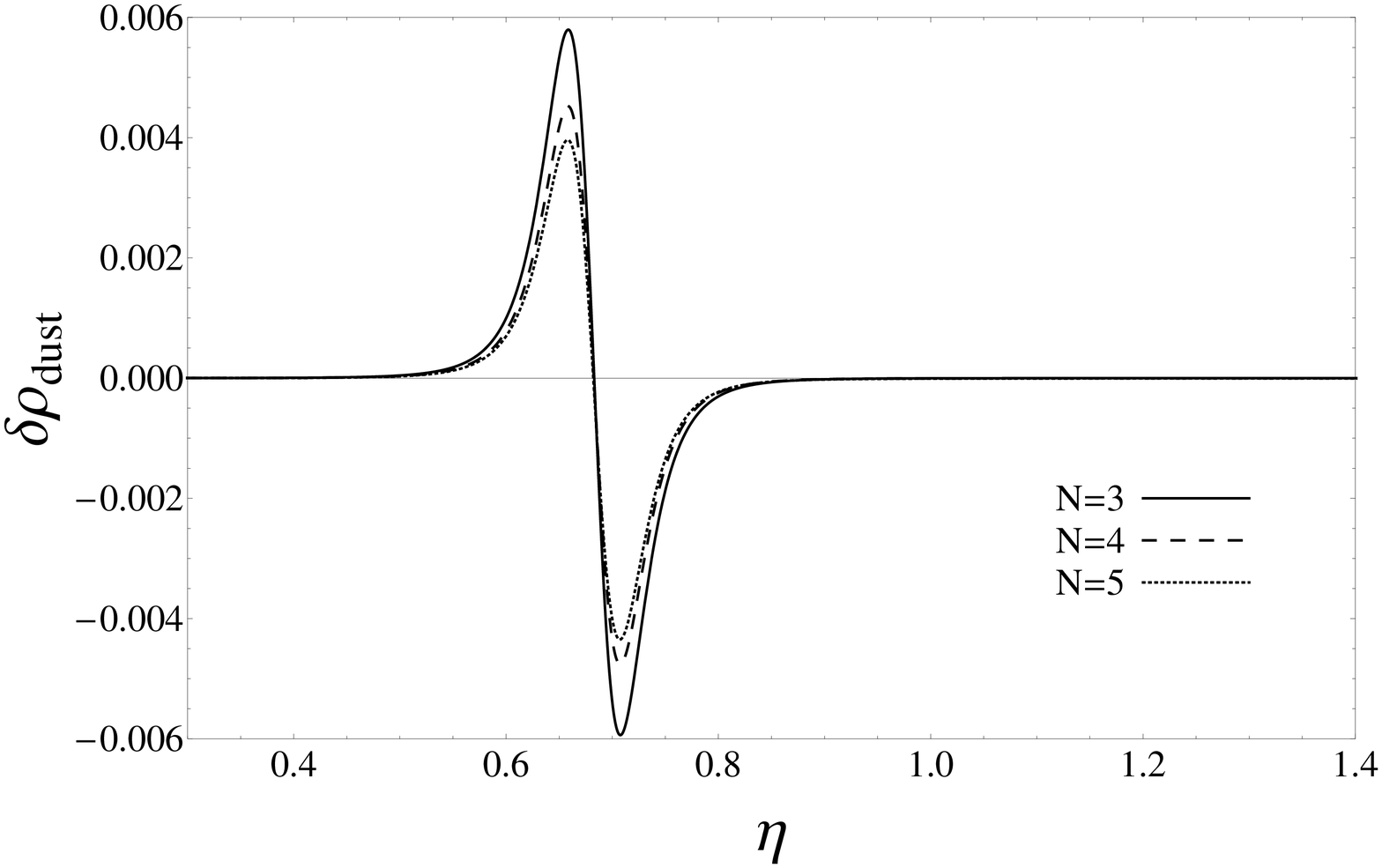}
\includegraphics*[height=5cm,width=8.5cm]{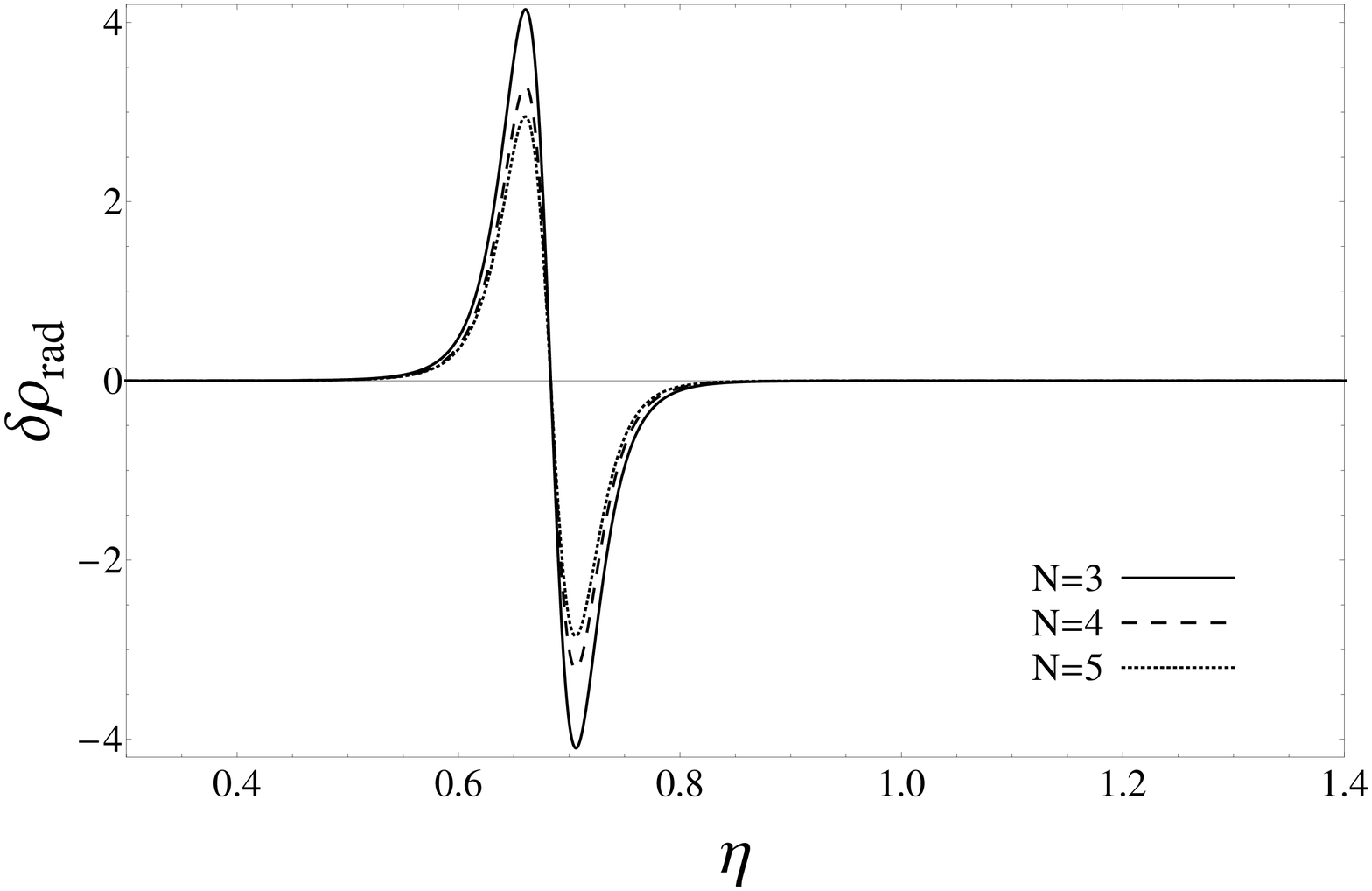}
\includegraphics*[height=5cm,width=8.5cm]{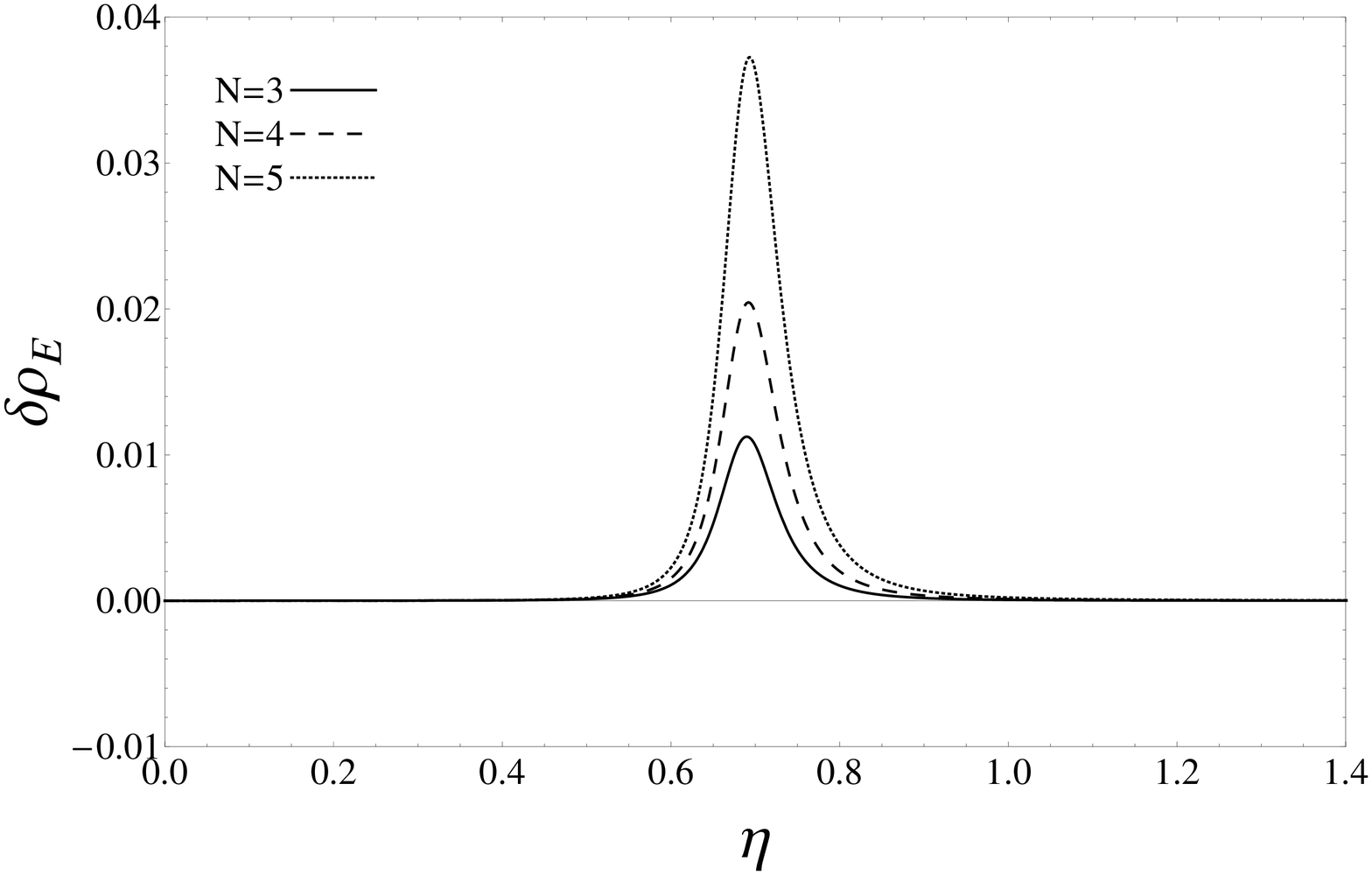}
\caption{Time evolution of the energy density fluctuations $\delta\rho_{\rm dust}$ (upper panel),
$\delta \rho_{\rm rad}$ (middle panel) and $\delta \rho_{E}$ (lower panel) for $N=3,4,5$ (solid, dashed and dotted curves,
respectively). In this case $|\delta \rho_{\rm dust} (\eta)|\simeq 10^{-2}$ and $|\delta \rho_{\rm rad}
(\eta)|\simeq 10$, with $|{\delta \rho_{\rm rad}}/{\rho_{\rm rad}}| \simeq |{\delta \rho_{\rm dust}}/{\rho_{\rm dust}}| \leq 10^{-
2}$
in a finite neighbourhood of the bounce.
As we increase $N$, the amplifications of the energy densities of dust and radiation enhanced by the
bounce in the respective mode decrease. The increase of $N$ enhances the amplifications on the Weyl energy density which may
diverge for
sufficiently smaller scales.}
\label{fig:deltarhoq}
\end{figure}

Fig.~\ref{fig:deltarhoq} (upper and middle panels) displays the behaviour of the energy density fluctuations for $N=3,4,5$. In
this
case $|\delta \rho_{\rm rad} (\eta)|\lesssim 10$ and $|\delta \rho_{\rm dust}(\eta)| \lesssim 10^{-2}$ in any
finite neighbourhood of the bounce, however we obtain

\begin{equation}\label{eqn:eqpin2new}
\left|\frac{\delta \rho_{\rm rad}}{\rho_{\rm rad}}\right|_{\Delta a}\simeq \left|\frac{\delta
\rho_{\rm dust}}{\rho_{\rm dust}}\right|_{\Delta a}
\leq 10^{-2}.
\end{equation}
As we increase $N$ (which is connected to the comoving scales of the perturbations)
the amplifications (in the respective mode) -- enhanced by the bounce -- of the
energy densities of dust and radiation decrease. On the contrary, the increase of $N$ increases
the amplifications on the Weyl energy density which may diverge for sufficiently smaller
scales (cf. the lower panel of Fig.~\ref{fig:deltarhoq}).

\begin{figure}
\includegraphics*[height=5cm,width=8.5cm]{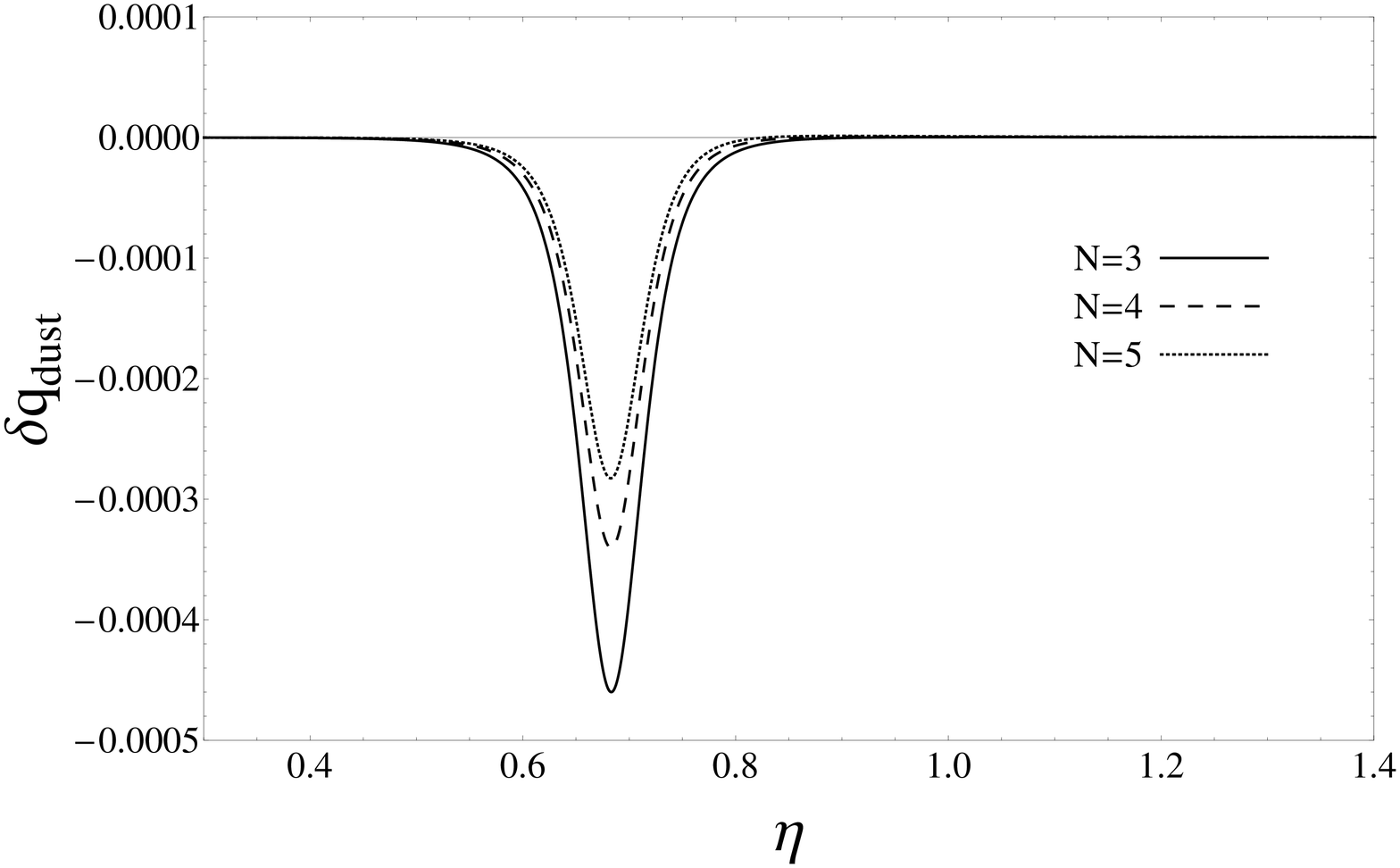}
\includegraphics*[height=5cm,width=8.5cm]{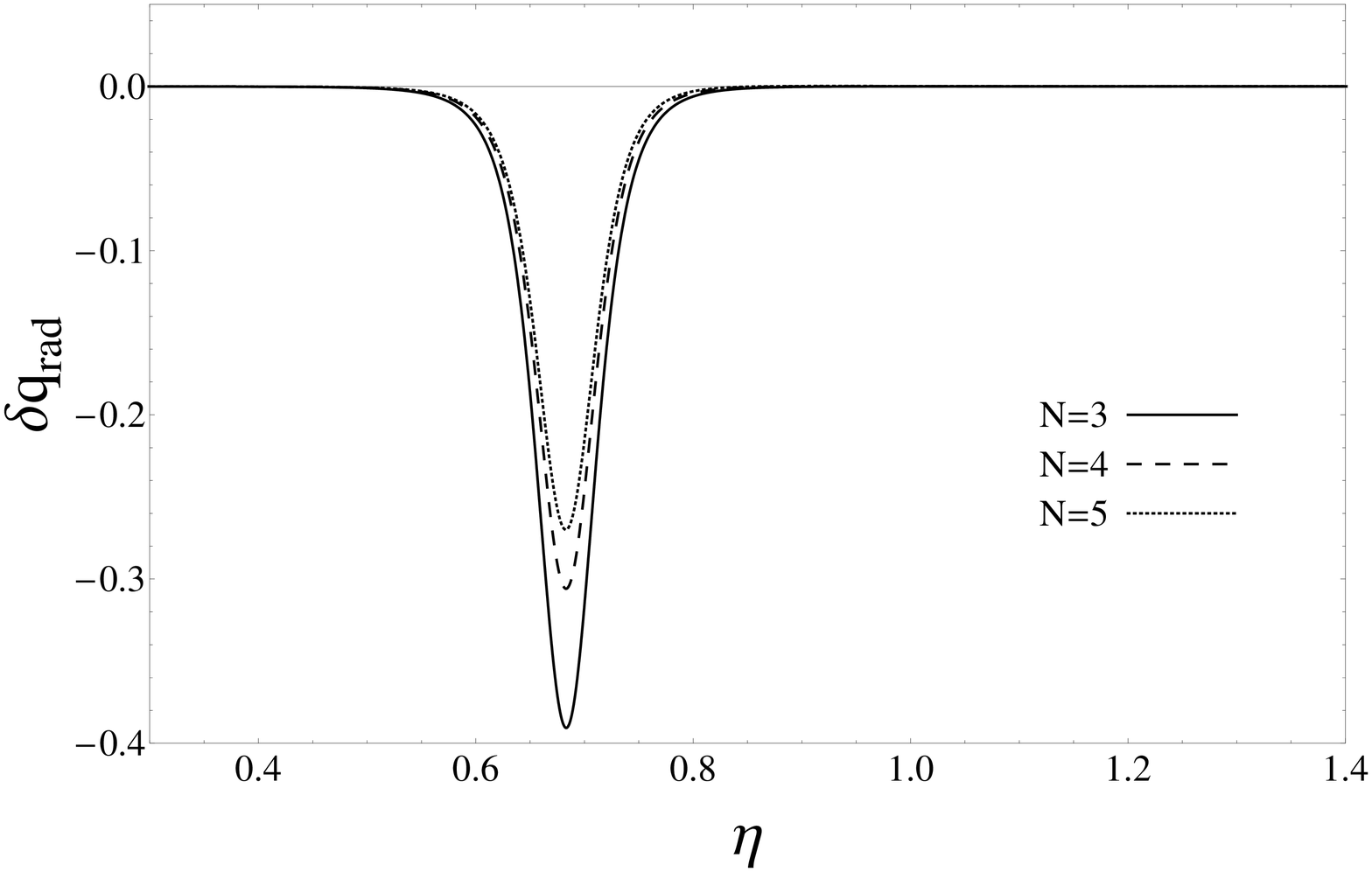}
\includegraphics*[height=5cm,width=8.5cm]{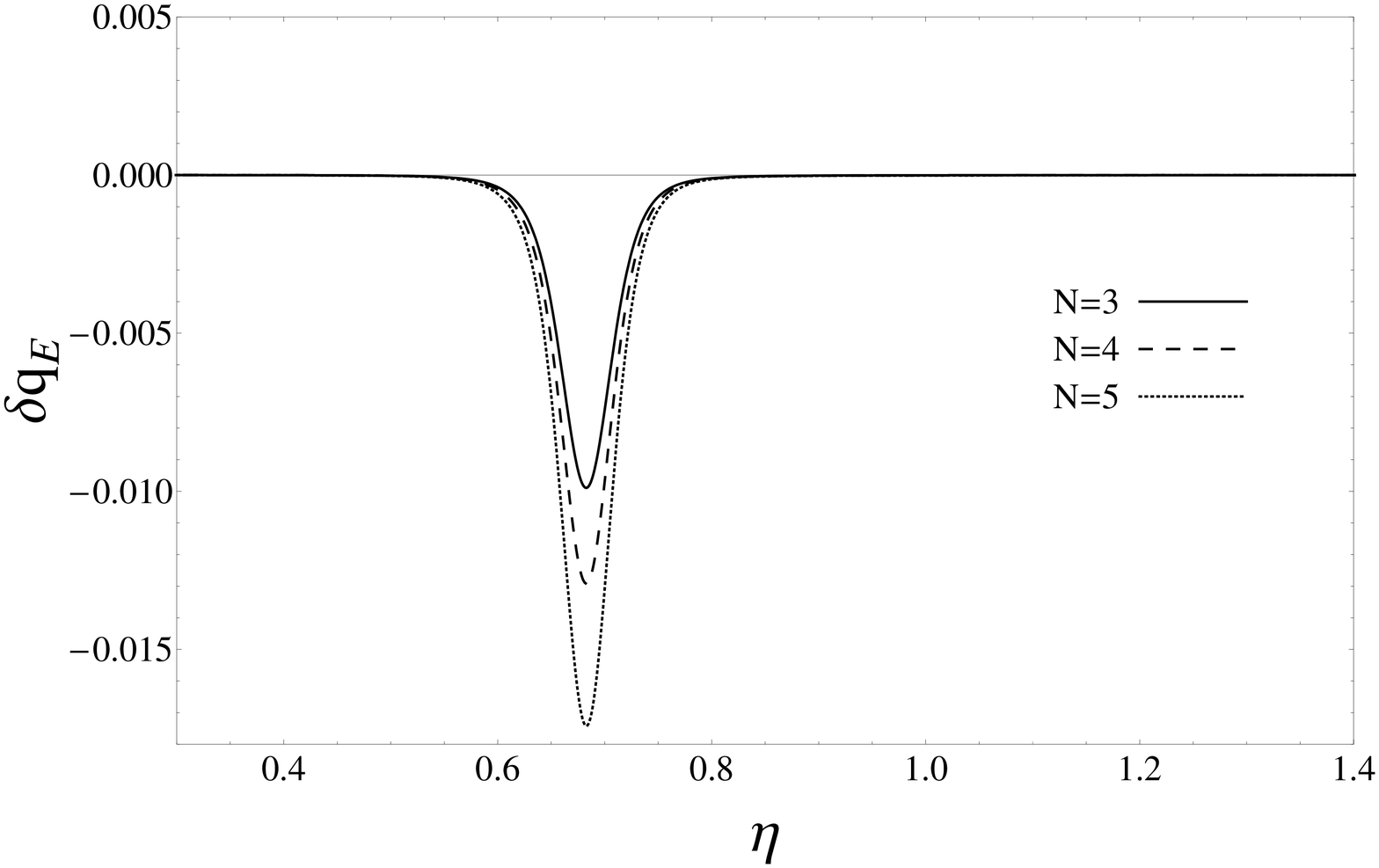}
\caption{Evolution of the heat fluxes perturbations $\delta q_{\rm dust}$ (upper panel), $\delta q_{\rm rad}$
(middle panel) $\delta q_{E}$ (lower panel) for $N=3,4,5$ (solid, dashed and dotted curves, respectively). The increase of $N$
makes the amplifications - enhanced by the bounce - of the heat fluxes of dust and radiation smaller. The increase of $N$
enhances the amplifications on the Weyl heat flux which may reach large values for sufficiently smaller scales.}
\label{fig:deltaqE}
\end{figure}

\begin{figure}
\includegraphics*[height=5cm,width=8.5cm]{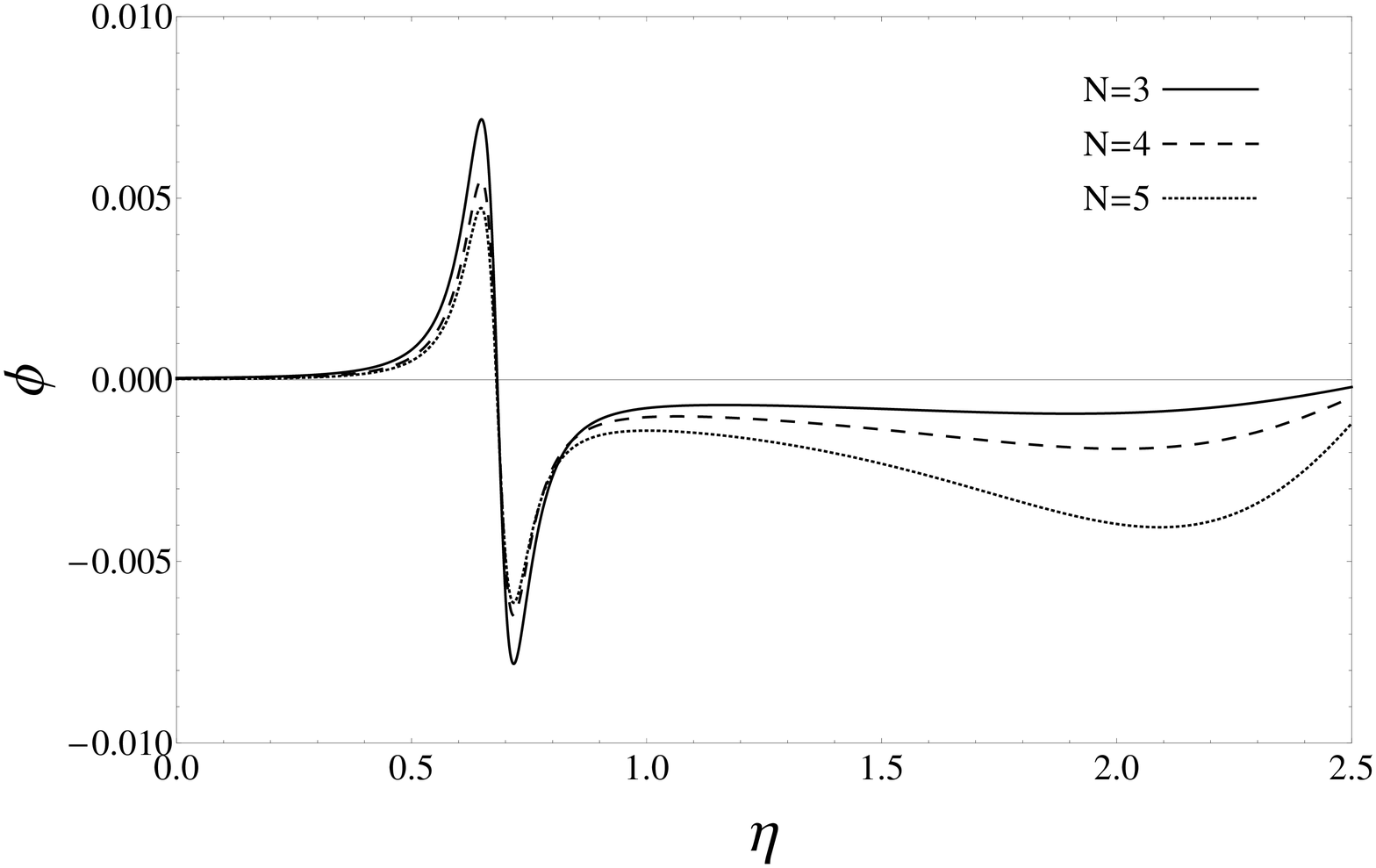}
\caption{Evolution of the potential $\phi$ in the case of a perturbed bulk for $N=3,4,5$ (solid, dashed and dotted curves,
respectively).
Opposite to the case of a de Sitter bulk, the perturbations remain bounded as the universe enters the late accelerated phase
and tend to zero as the universe approaches the final de Sitter state, for any $N$.}
\label{fig:phinewer}
\end{figure}
\par
In Fig.~\ref{fig:deltaqE} we show the evolution of the heat fluxes perturbations. Again,
while the amplifications of the heat fluxes of dust and radiation decrease with the increase of $N$,
the heat flux of the Weyl fluid tends to increase for smaller scale perturbations.

\par However a striking difference (cf. Figs.~\ref{fig:phinew} and \ref{fig:phinewer}) exists
between the dynamics of the perturbations for a de Sitter bulk
and for a perturbed bulk. In the case of a de Sitter bulk all the perturbation variables, for any $N$, increase as
the universe enters the late accelerated phase, diverging as it approaches the final de Sitter state. On the contrary,
in the case of a perturbed bulk, all the perturbations remain bounded as the universe enters the late accelerated phase
and tend to zero as the universe approaches the final de Sitter state, for any $N$. In this sense we consider that
linear perturbations, for one bounce orbits in region II (cf. Fig.~\ref{fig:phasespace}), are dynamically more stable in the case
of a perturbed de Sitter bulk.

Apart from the energy density of the Weyl component,
the amplitudes of all perturbations remain sufficiently small and bounded
relative to the background values for
any scale of the perturbations. In fact it can be numerically shown that $|\phi|\simeq 10^{-3}$ in a finite neighbourhood of the
bounce
for the case $N=15$. Furthermore, the dominant background hydrodynamical fluid component in the bounce is $\rho_{\rm rad}\simeq
200$.
Comparing this quantity to the amplitude growth of the Weyl energy density perturbation in the bounce, we obtain
\begin{equation}\label{eqn:eqpin2new}
\left|\frac{\delta \rho_{\rm E}}{\rho_{\rm rad}}\right|_{\Delta a ~(N=15)}\simeq 10^{-2}~.
\end{equation}
Thus, contrary to the case of of a de Sitter bulk, the linear perturbative regime still holds
for $N=15$ in a perturbed bulk. In this sense, a perturbed bulk is ``more stable'' when compared to a de Sitter bulk.

\section{Eternal Universes}\label{sect:EUPBulk}

From a theoretical point of view it is also interesting to examine how the bounce affects the amplitude growth of
of scalar perturbations in orbits of region I (see Fig.~\ref{fig:phasespace}). These orbits describe
perpetually bouncing universes so that this may shed some light on the treatment of linear perturbations in the framework of
eternal bouncing
models. Since we have already seen that the presence of one bounce amplifies the linear scalar hydrodynamical perturbations,
we would expect that the successive infinite bounces -- in orbits of Region I -- would produce
recurring amplifications which would soon violate the linear approximation breaking the stability of eternal universes.
\par
In order to simplify our analysis, we will restrict ourselves again to the cases $N=3,4,5$.
From the dynamical point of view, the enhancement in the amplitudes of scalar perturbations due to the bounce
should be qualitatively the same for smaller scales.

\par
\begin{figure}
\includegraphics*[height=5cm,width=8cm]{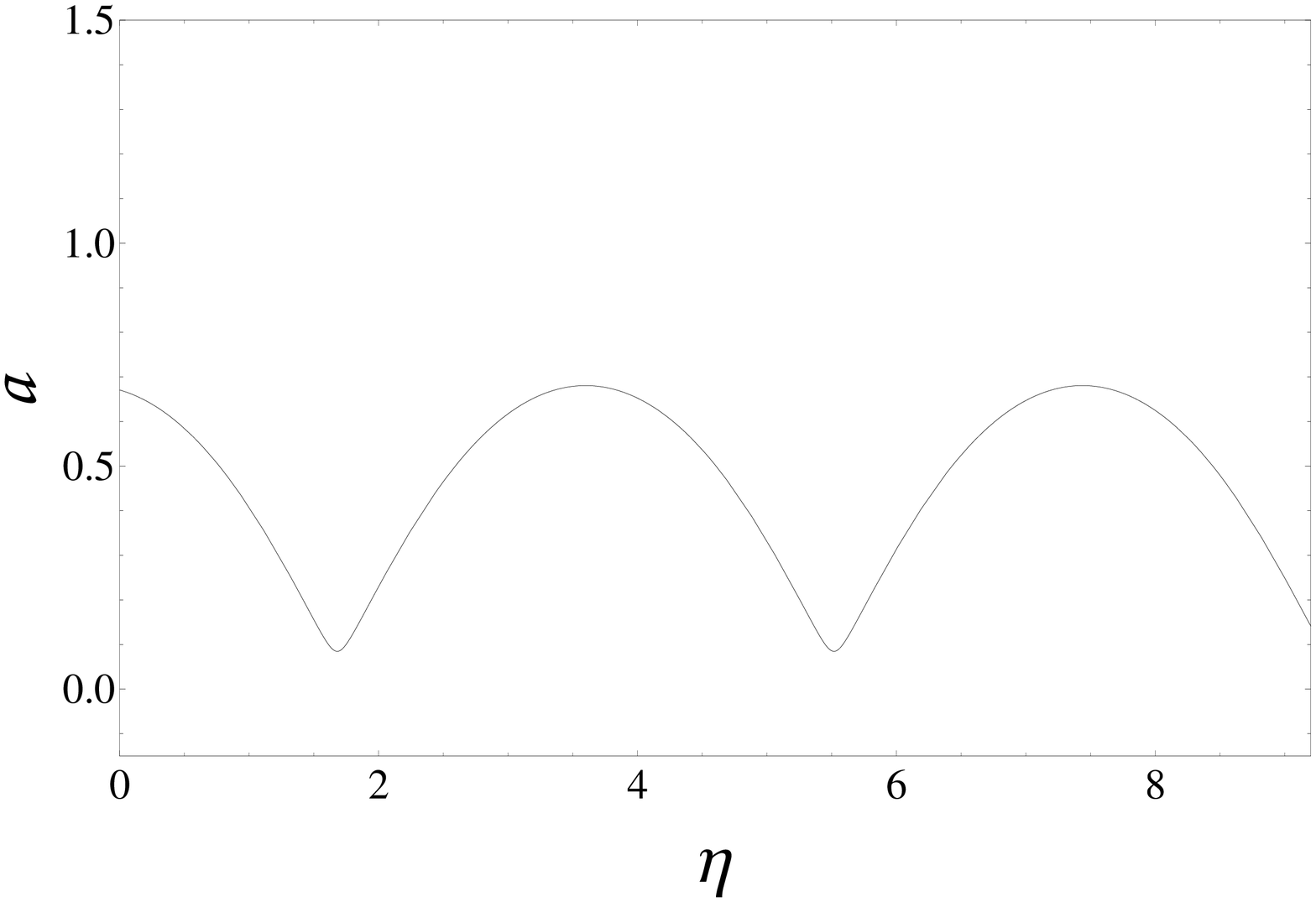}
\caption{The scale factor of an eternal universe of Region I. Here we fix the parameters
$\kappa^2_4=100$, $|\sigma|=100$, $E_{\rm dust}=0.001$,
$E_{\rm rad}=0.01$, $\Lambda=1.5$,
and the initial condition
$p_{a_0}=-0.5$. The exact $16$ digit
precision values for the scale factor initial condition is given by $a_0=0.670686673645725$
so that (\ref{eqn:eq4}) is satisfied.
For $\eta\lesssim8$ there are two different values
of $\bar{\eta}_b$ connected to two bounces. The scale of the bounces is $a(\bar{\eta}_b)\simeq 0.1$.}
\label{fig:a_etaEU}
\end{figure}

To numerically solve the equations for scalar perturbations, let us consider a given orbit in region I.
It can be shown that one of those orbits is generated by fixing the parameters and initial conditions
given in Fig.~\ref{fig:a_etaEU}.
In this case the scale of the bounce is $a(\bar{\eta}_b)\simeq 0.1$ (see Fig.~\ref{fig:a_etaEU}).
This will be the background configuration in Region I that we now impose in order to examine the behaviour of
scalar perturbations in a de Sitter or a perturbed bulk.

\subsection{The Evolution of Scalar Perturbations in a FLRW Brane Embedded in a de Sitter Bulk}

\par
\begin{figure}
\includegraphics*[height=4.8cm,width=7.9cm]{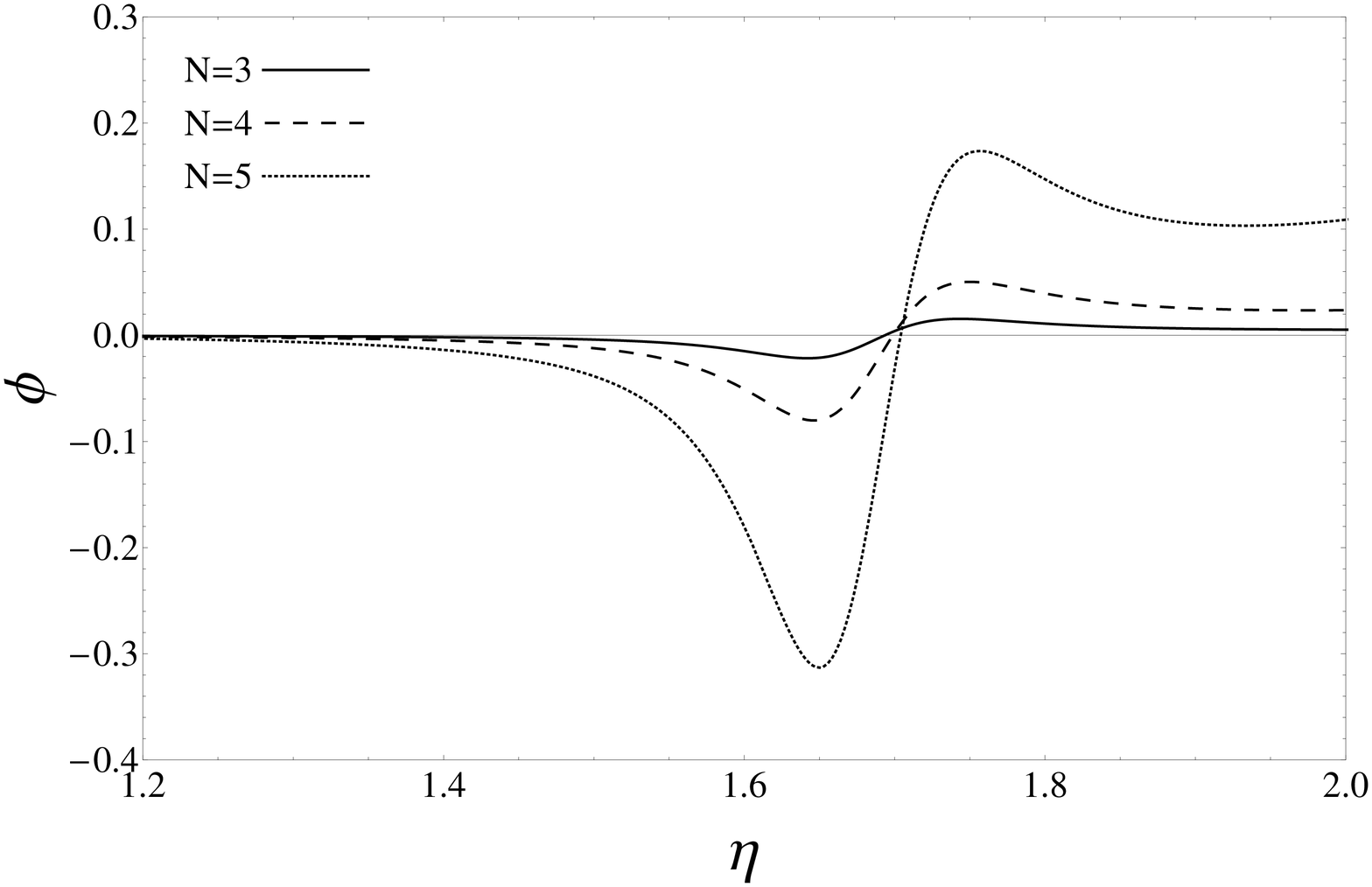}
\includegraphics*[height=4.8cm,width=7.9cm]{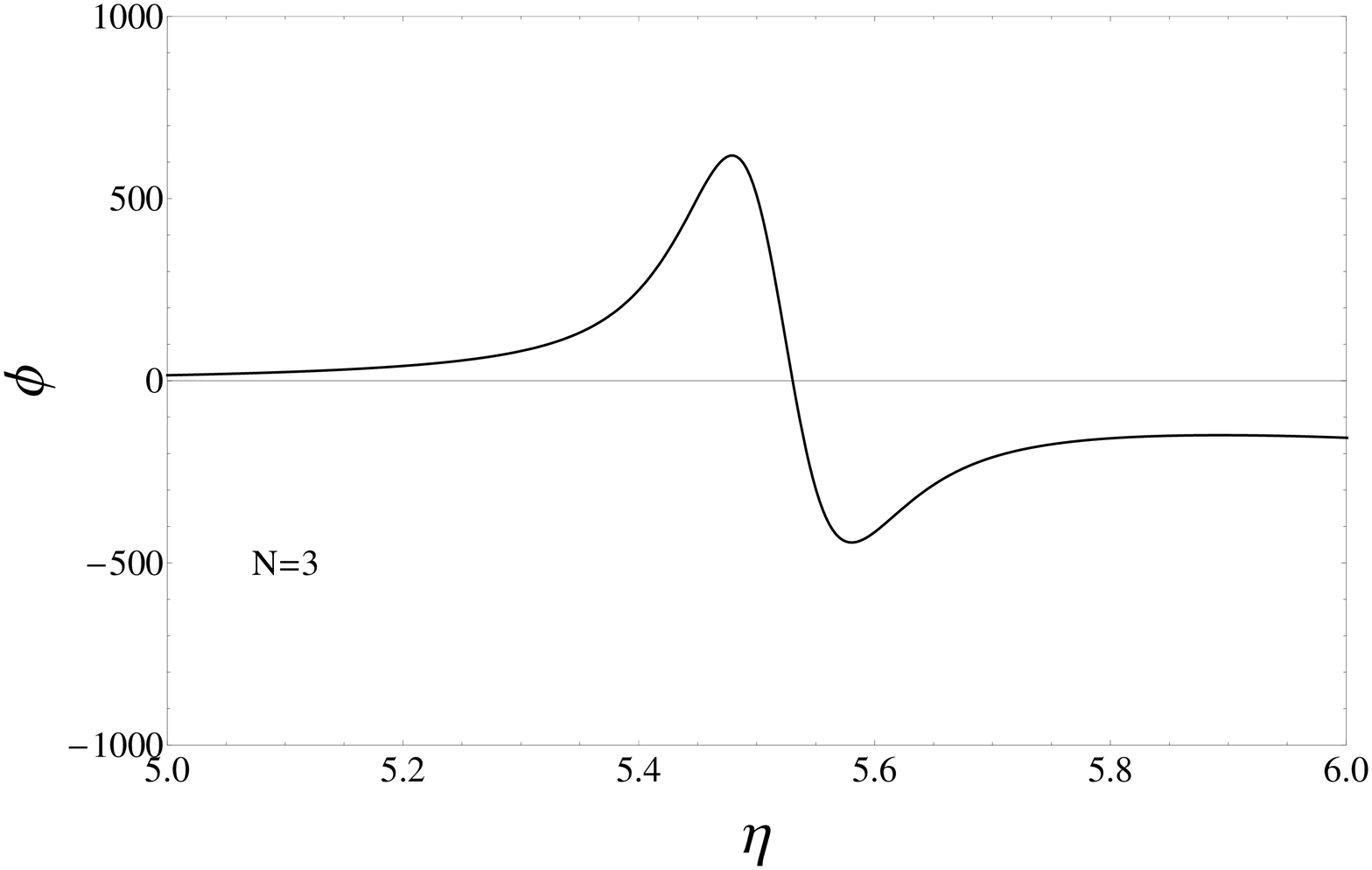}
\caption{The behaviour of the potential $\phi$ for $N=3,4,5$ in a finite neighbourhood first bounce (upper panel), and for $N=3$
in a finite neighbourhood second bounce (lower panel). Here we fix
$\delta\rho_{0~\rm rad}=10^{-5}$ and
$\delta\rho_{0~\rm dust}=\delta q_{0~\rm dust}=\delta q_{0~\rm rad}=0$
as the initial conditions for the hydrodynamical perturbations. The exact $16$ digit
precision values for the initial condition of $\psi$ is given by $\psi_0=0.000224791696635$
so that (\ref{eqn:eqv2}) together with (\ref{eqn:eq2n}) is satisfied. It is worth remarking that $|\phi(\eta)|_{N=4,5,..} >> |\
phi(\eta)|_{N=3}$, in a finite neighbourhood of the second bounce $\Delta a=|a(\eta)-a({\eta}_b)|$.
}
\label{fig:phipsiEU}
\end{figure}

 Let us first consider eternal universes in the case of a de Sitter bulk.
In Fig.~\ref{fig:phipsiEU} we show the behaviour of the potential $\phi$ in a finite neighbourhood of the first two bounces
adopting a particular set of initial conditions for the hydrodynamical perturbations.
In a domain of $\eta \leq 8$, which includes 2 bounces at $a(\bar{\eta}_b)\simeq 0.1$ each, we
can see that
$|\phi(\eta)|\sim 10^{-1}$ for $N=3,4,5$ in a finite neighbourhood of the first bounce $\Delta a=|a(\eta)-a({\eta}_b)|$,
while
$|\phi(\eta)|\sim 10^{3}$ for $N=3$ in a finite neighbourhood of the second bounce $\Delta a=|a(\eta)-a({\eta}_b)|$.
This suggests that the first bounce does not spoil
the scalar perturbations connected to hydrodynamical fluctuations while the second one does.
\begin{figure}
\includegraphics*[height=5cm,width=8.5cm]{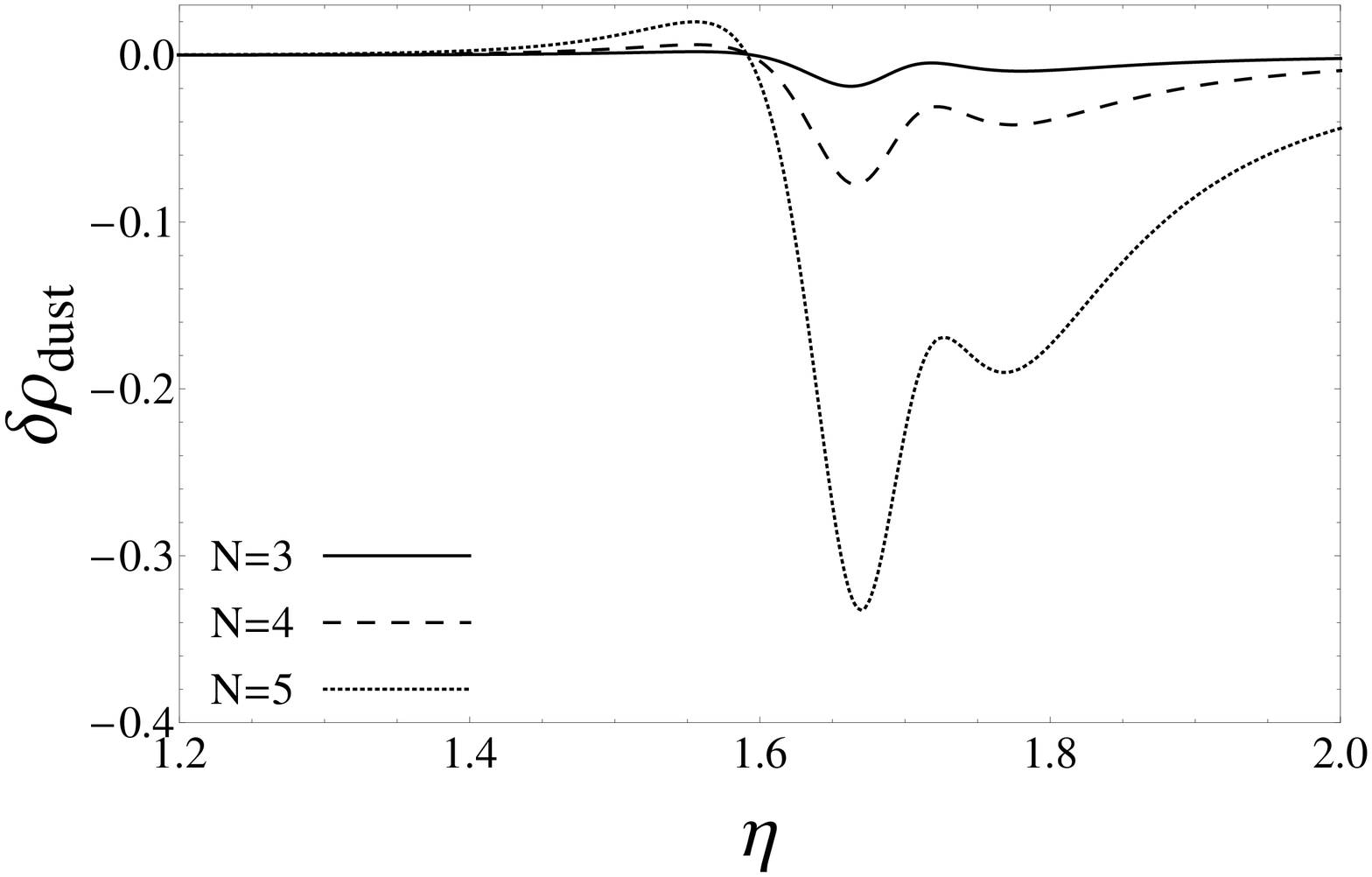}
\includegraphics*[height=5cm,width=8.5cm]{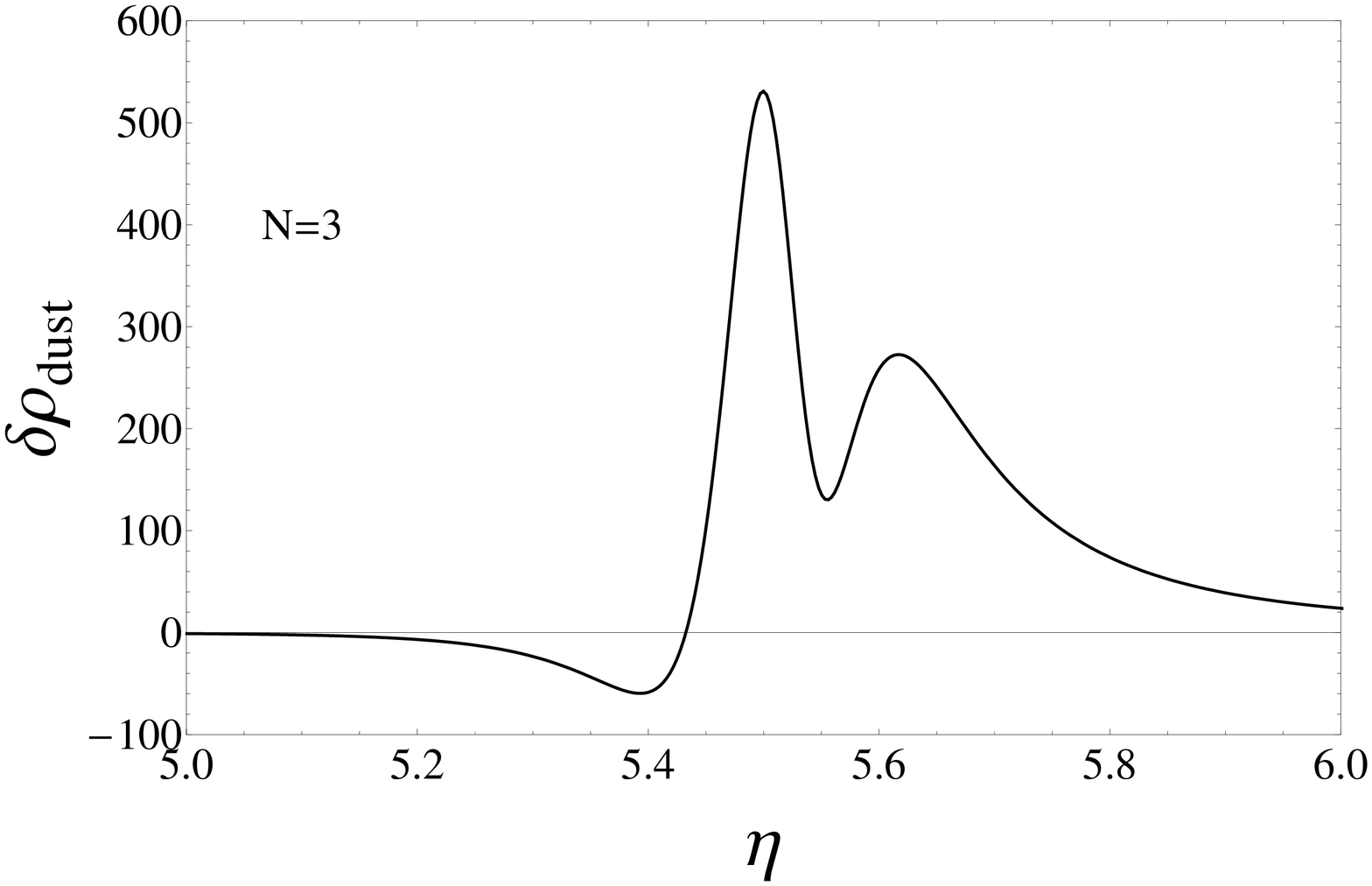}
\caption{Time evolution of the energy density fluctuations $\delta\rho_{\rm dust}$ for $N=3,4,5$ in a finite neighbourhood first
bounce (upper panel), and
for $N=3$ in a finite neighbourhood of the second bounce (lower panel).
In this case $|\delta \rho_{\rm dust} (\eta)|\simeq 10^{-1}$ in a finite neighbourhood of the first bounce,
while $|\delta \rho_{\rm dust} (\eta)|\simeq 10^{3}$ in a finite neighbourhood of the second bounce.
The linear regime is broken already in the second bounce, at least.
This illustrates how eternal universes are highly unstable configurations when a de Sitter bulk is taken into account.}
\label{fig:phiEU1}
\end{figure}
\par In fact, according to Fig.~\ref{fig:phiEU1} the amplitudes of energy density fluctuations (for $N=3,4,5$) is of the
order $|\delta\rho_{dust}|\sim0.1$ in a finite neighbourhood of the first bounce $\Delta a=|a(\eta)-a({\eta}_b)|$.
Thus,
\begin{equation}\label{eqn:eqpin2newr}
\left|\frac{\delta\rho_{\rm dust}}{\rho_{\rm dust}}\right|_{\Delta a, N=3,4,5}\lesssim 10^{-1}.
\end{equation}
On the other hand in the second bounce the amplitude of energy density fluctuations (for $N=3$) is of the
order $|\delta\rho_{dust}|\sim1000$ in a finite neighbourhood of the second bounce $\Delta a=|a(\eta)-a({\eta}_b)|$.
Thus,
\begin{equation}\label{eqn:eqpin2newr2}
\left|\frac{\delta\rho_{\rm dust}}{\rho_{\rm dust}}\right|_{\Delta a, N=3}\lesssim 10^{3},
\end{equation}
breaking the linear regime. As we increase $N$ (which is connected to the comoving scales of the perturbations)
the amplifications (in the respective mode) -- enhanced by the second bounce -- of the
energy densities fluctuations also increase.
\subsection{The Evolution of Scalar Perturbations in a FLRW Brane Embedded in a Perturbed Bulk}
\par
\begin{figure}
\includegraphics*[height=4.8cm,width=7.9cm]{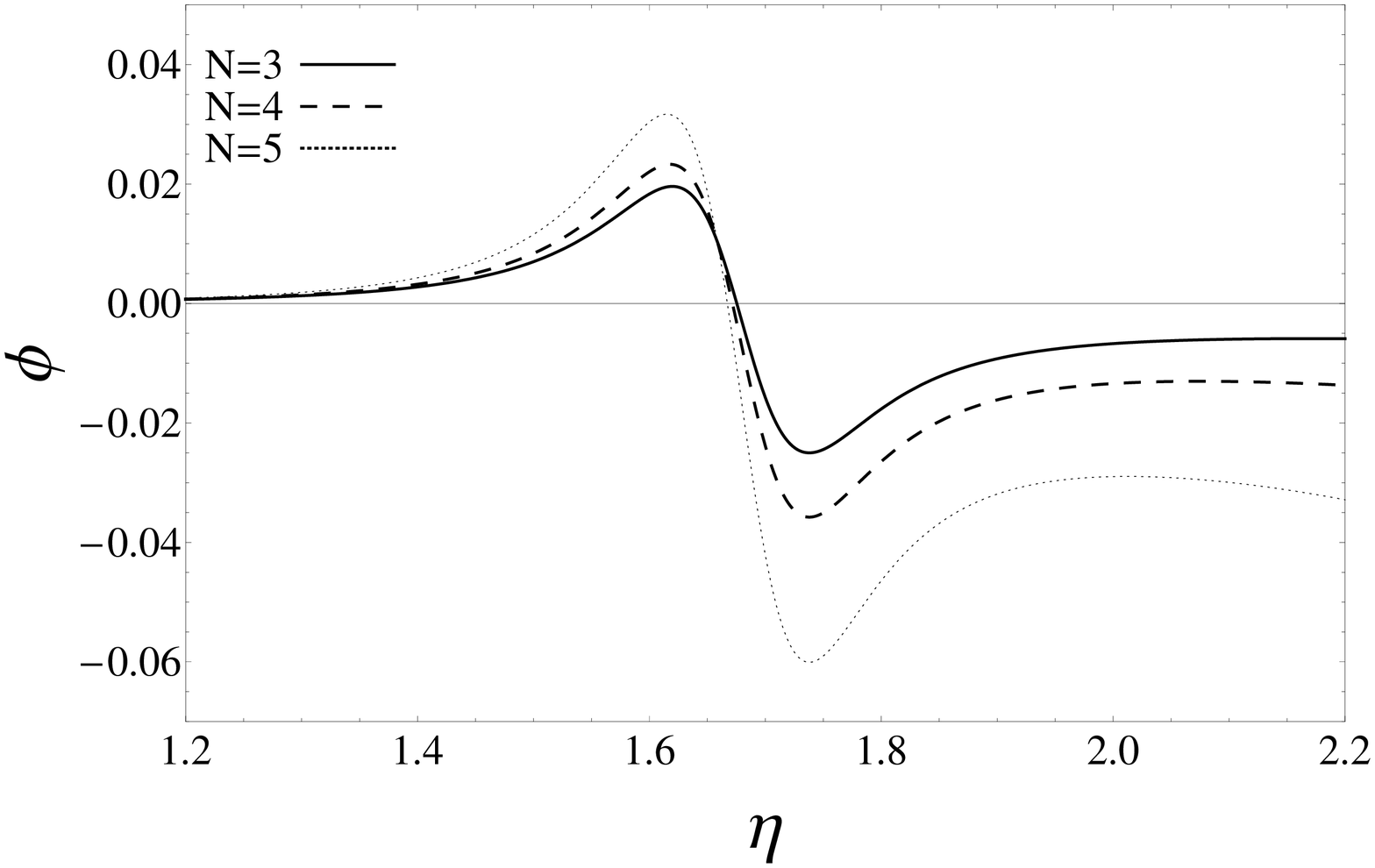}
\includegraphics*[height=4.8cm,width=7.9cm]{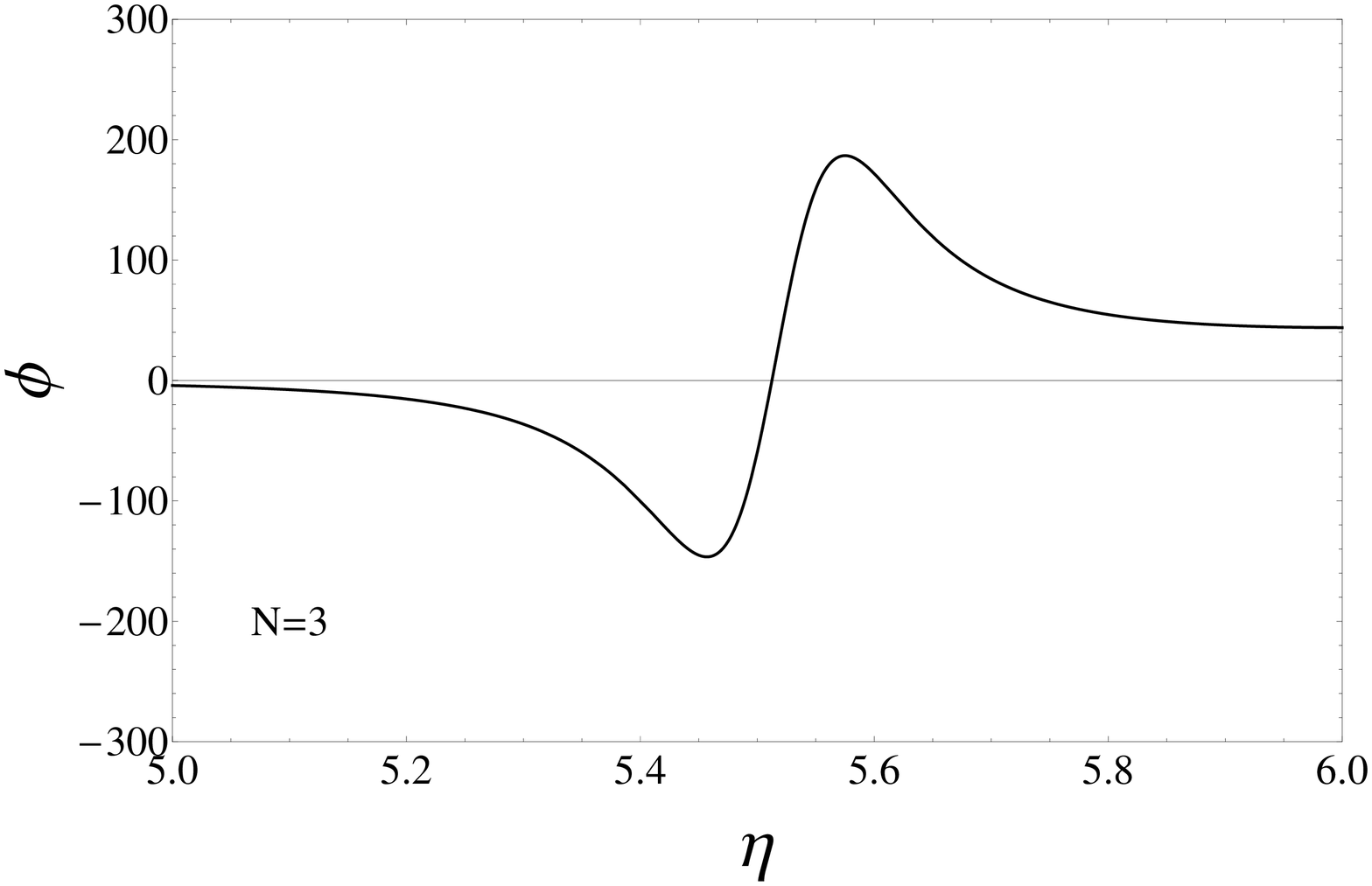}
\caption{The behaviour of the potential $\phi$ for $N=3,4,5$ in a finite neighbourhood first bounce (upper panel), and for $N=3$
in a finite neighbourhood second bounce (lower panel). Here we fix
$\delta\rho_{0~\rm rad}=10^{-5}$, $\delta \rho_{E}(0)=10^{-6}$ and
$\delta\rho_{0~\rm dust}=\delta q_{0~\rm dust}=\delta q_{0~\rm rad}=\delta q_{E}(0)=0$
as the initial conditions for the hydrodynamical perturbations. The exact $16$ digit
precision values for the initial condition of $\psi$ is given by $\psi_0=0.000224791696635$
so that (\ref{eqn:pb3}) is satisfied. It is worth remarking that $|\phi(\eta)|_{N=4,5,..} >> |\phi(\eta)|_{N=3}$, in a finite
neighbourhood of the second bounce $\Delta a=|a(\eta)-a({\eta}_b)|$.}
\label{fig:phiEU2}
\end{figure}

Let us now consider eternal universes in the case of a perturbed bulk.
In Fig.~\ref{fig:phiEU2} we show the behaviour of the potential $\phi$ in a finite neighbourhood of the first two bounces
adopting a particular set of initial conditions for the hydrodynamical perturbations.
For a time ${\eta}\leq 2$, we notice that $|\phi({\eta})| \sim 10^{-2}$ in a finite neighbourhood of the first bounce so that the
large scale amplifications (namely, $N=3,4,5$) in the potentials about a neighbourhood of each of the first bounces
remain sufficiently bounded. In this sense, the first bounce does not spoil the scalar metric perturbations.
On the other hand, for a time $6\leq{\eta}\leq 5$, $|\phi({\eta})| \sim 10^{2}$ in a finite neighbourhood of the second bounce.
Comparing with Fig.~\ref{fig:phipsiEU} we see that although the presence of the Weyl fluid reduces the growth rate of the
potential
$\phi$, the second bounce already violates the linear regime as in the case of a de Sitter bulk.

\begin{figure}
\includegraphics*[height=5cm,width=8.5cm]{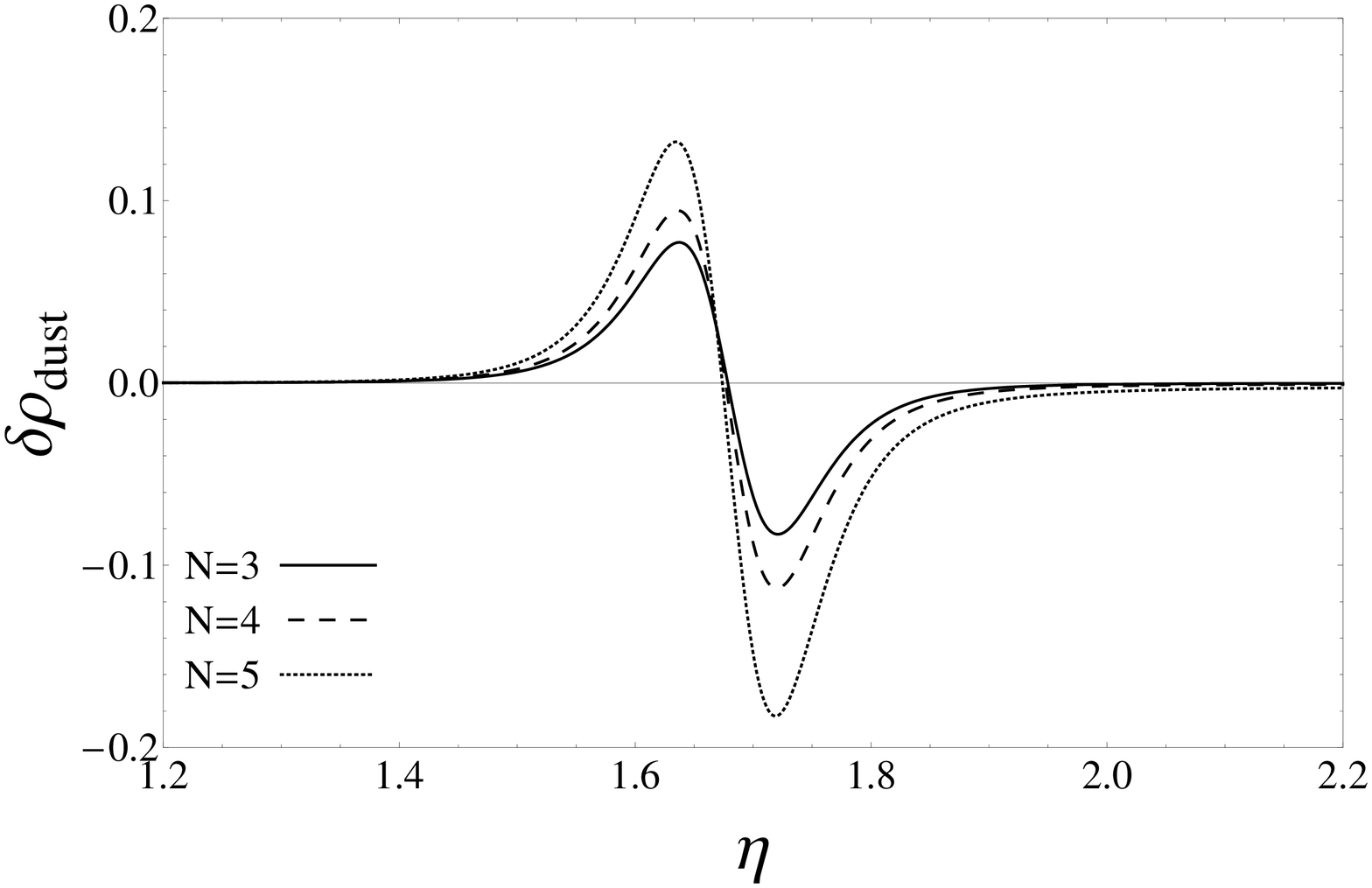}
\includegraphics*[height=5cm,width=8.5cm]{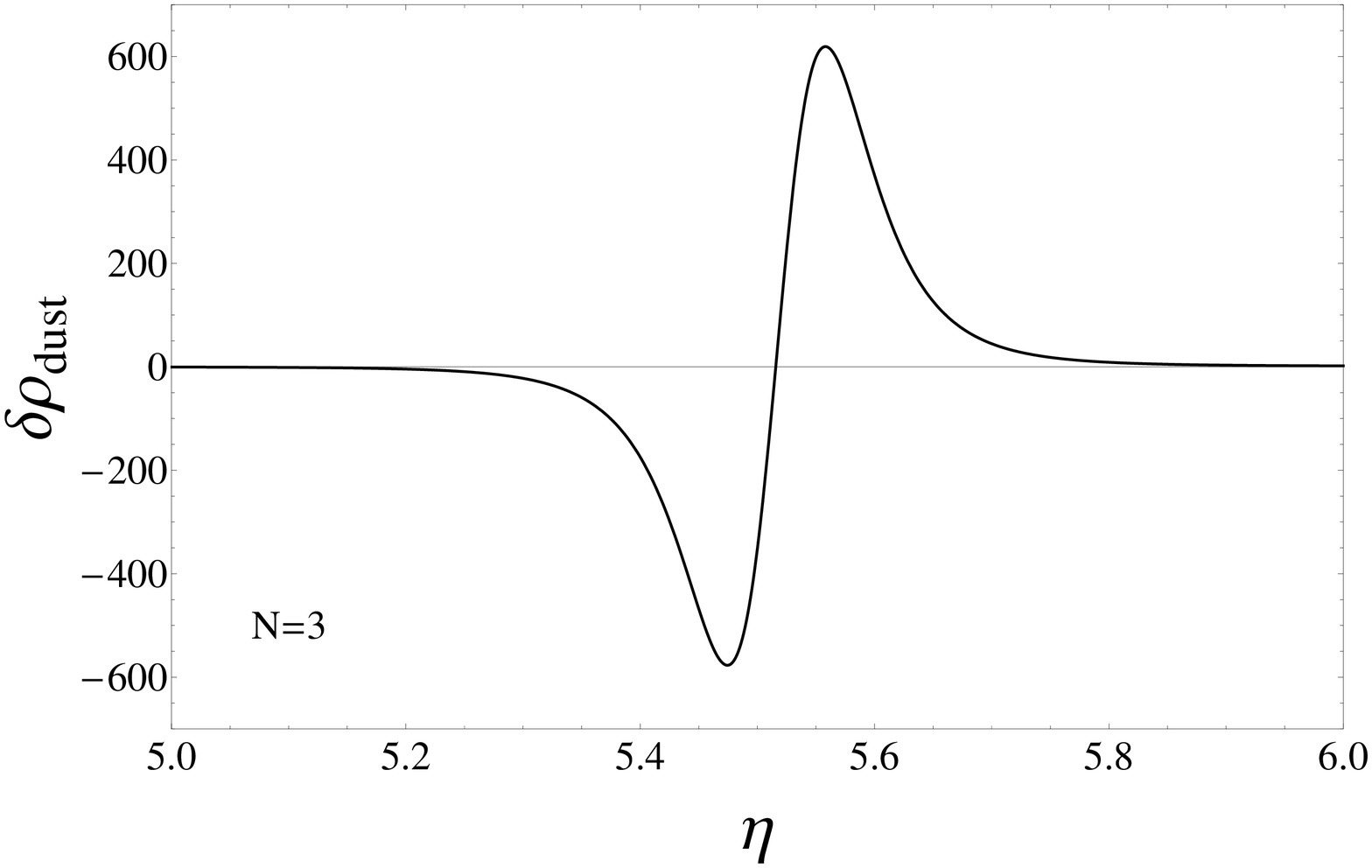}
\caption{Time evolution of the energy density fluctuations $\delta\rho_{\rm dust}$ for $N=3,4,5$ in a finite neighbourhood first
bounce (upper panel), and
for $N=3$ in a finite neighbourhood of the second bounce (lower panel).
In this case $|\delta \rho_{\rm dust} (\eta)|\simeq 10^{-1}$ in a finite neighbourhood of the first bounce,
while $|\delta \rho_{\rm dust} (\eta)|\simeq 10^{2}$ in a finite neighbourhood of the second bounce.
The linear regime is again broken in the second bounce.
This illustrates how eternal universes are also highly unstable configurations when a perturbed bulk is taken into account.}
\label{fig:phideltarho}
\end{figure}

According to Fig.~\ref{fig:phideltarho} the amplitudes of energy density fluctuations (for $N=3,4,5$) is of the
order $|\delta\rho_{dust}|\sim0.1$ in a finite neighbourhood of the first bounce $\Delta a=|a(\eta)-a({\eta}_b)|$.
Thus, again we obtain $\delta\rho_{\rm dust}/{\rho_{\rm dust}}|_{\Delta a, N=3,4,5}\lesssim 10^{-1}$.
On the other hand, in the second bounce the amplitude of energy density fluctuations (for $N=3$) is of the
order $|\delta\rho_{dust}|\sim 10^{2}$ in a finite neighbourhood of the second bounce $\Delta a=|a(\eta)-a({\eta}_b)|$, so that ${
\delta\rho_{\rm dust}}/{\rho_{\rm dust}}|_{\Delta a, N=3}\sim 10^{2}$. Again, the linear regime is broken at least in the second
bounce.
As we increase $N$,
the amplifications of
energy densities fluctuations also increase substantially in the second bounce.
\section{Conclusions and Final Discussion}\label{sect:discussion}

In this work we have studied the dynamics and the evolution of scalar hydrodynamical perturbations in closed FLRW
universes within the framework of a brane world theory with an extra timelike dimension.
In brane world cosmologies, classical General Relativity is recovered in the low-energy limit, while different effects
appear in the high-energy limit: in particular, in the case of a timelike extra dimension, it is possible to avoid the
initial singularity that affects the usual cosmological models based on General Relativity.
We considered some background models, solutions of the modified Einstein field equations in the brane, with a
cosmological constant term and source terms given by two perfect fluids, dust and radiation.
The bulk-brane interaction modifies the Einstein equations introducing extra terms in the dynamics
which avoid the emergence of a singularity. For a closed FLRW geometry the phase-space of the background
presents only two critical points, a centre (associated with the minimum of the potential) and a saddle
(associated with the local maximum of the potential) corresponding
to Einstein universe configurations, stable and unstable, respectively. The phase space is separated
into two regions, Region I (around the centre) of closed orbits describing eternal universes and Region II (outside the
separatrices
originating from the saddle) of open orbits with one bounce\cite{maier}.
\par The core of this paper is a numerical analysis of scalar perturbations which can shed some light on the study of
cosmological scalar perturbations in bouncing brane world models. This analysis is done separately for one bounce universes and
eternal universes (with
infinitely many bounces). Also two separate classes of perturbation dynamics are considered, namely,
one with a de Sitter bulk and the other with a perturbed bulk.
\par
Although the orbits in region III are mathematically sound one bounce solutions,
they were no considered in our analysis since they do not provide a concrete cosmological bouncing model
connected with observations, as the orbits in region II of Fig.~\ref{fig:phasespace} do\cite{mns}.
Motivated by this observational feature, we restricted ourselves
to one bounce orbits in region II (instead of region III).
According to our numerical analysis, although the bounce enhances the amplitudes of
scalar perturbations for these class of one bounce solutions in the case of a de Sitter bulk,
the amplitudes of the perturbations remain sufficiently small and bounded
relative to the background values up to a certain scale, namely $N=15$.
For a perturbed bulk, the amplitudes of all perturbations -- apart from the energy density of the Weyl component -- remain
sufficiently small and bounded
relative to the background values for any scale of the perturbations.
Furthermore, for a perturbed bulk, all the perturbations
in the case of one bounce universes remain bounded as the universe enters the late accelerated phase
and tend to zero as the universe approaches the final de Sitter state, for any comoving scale $N$.
This behaviour is opposite to the case of a de Sitter bulk where all variables rapidly increase as
the universe enters the late accelerated phase, diverging as it approaches the final de Sitter state,
for any comoving scale $N$. In this sense we consider that linear hydrodynamical perturbations,
for these class of one bounce models are dynamically more stable in the case of a perturbed bulk.
\par A careful numerical analysis shows us that eternal universes are, from
the numerical point of view, highly unstable configurations. In fact, if we consider scalar
perturbations in a given orbit with initial conditions in a neighbourhood of
the stable static Einstein universe (the critical point $P_0$, cf. Fig. 3), we can see that although the constraints
(\ref{eqn:eqsp7new}) (for a de Sitter bulk) or (\ref{eqn:pb3}) (for a perturbed bulk) are preserved,
the amplifications of hydrodynamical perturbations are too large -- already at early times, typically in the second bounce --
when compared to the background values.
Since the constraints (\ref{eqn:eqsp7new}) (for a de Sitter bulk) or (\ref{eqn:pb3}) (for a perturbed bulk)
are satisfied with an error less than $10^{-10}$, this is clearly
a dynamical instability which breaks the linear perturbative regime.
\par As a future perspective we intend to examine the role of the bounce -- with respect to scalar perturbations --
in the case that the observed cosmological parameters are taken into account\cite{mns}. For this case, the generalization of the
Mukhanov-Sasaki equation\cite{mukhanov}
might be a powerful tool in order to simplify the dynamical equations -- and hence their numerical solution -- that govern the
evolution of scalar perturbations.
In this case, we also intend to implement a better code in which we could increase values of $N$ together with the
current Planck constraints on the spatial curvature.

\section*{Acknowledgements}
RM and IDS acknowledge the financial support
from CNPq/MCTI-Brasil, through a post-doctoral research grant no. 201907/2011-9 (RM) and research grant
no. 306527/2009-0 (IDS). FP is supported by STFC grant ST/H002774/1.

\end{document}